\documentclass[conference]{IEEEtran}
\usepackage[english]{babel}
\usepackage{blindtext}
\usepackage{hyperref}
\hypersetup{
    colorlinks=true,
    linkcolor=blue,
    filecolor=magenta,      
    urlcolor=cyan,
}
\usepackage[T1]{fontenc}
\usepackage[utf8]{inputenc}
\usepackage{framed}
\usepackage{xurl}
\usepackage{xcolor}
\usepackage{wasysym}
\usepackage{booktabs}
\usepackage{graphicx}
\usepackage{paralist}
\usepackage[font=small,labelfont=bf]{caption}
\usepackage{subfigure}
\usepackage{stackengine}
\usepackage{pifont}
\usepackage[hang,flushmargin]{footmisc}
\renewcommand{\footnotesize}{\fontsize{8}{9}\selectfont}
\usepackage{enumitem}
\usepackage{ifthen}
\usepackage{cellspace}
\usepackage{multirow}

\usepackage[square,sort,comma,numbers]{natbib}

\usepackage{mdframed}

\mdfdefinestyle{MyFrame}{%
    linecolor=black,
    outerlinewidth=2pt,
    innertopmargin=2pt,
    innerbottommargin=2pt,
    innerrightmargin=2pt,
    innerleftmargin=2pt,
    leftmargin = 2pt,
    rightmargin = 2pt,
    backgroundcolor=gray!20
}

\definecolor{linkcol}{rgb}{0.3,0,0}
\definecolor{citecol}{rgb}{0.3,0,0}
\definecolor{urlcol}{rgb}{0.3,0,0}

\usepackage{xspace}
\usepackage{comment}

\newcommand{\descr}[1]{\smallskip\noindent\textbf{#1}}

\makeatletter
\def\url@leostyle{%
  \@ifundefined{selectfont}{\def\UrlFont{}}%
  {\def\UrlFont{}}%
}

\def\ps@IEEEtitlepagestyle{%
  \def\@oddfoot{\mycopyrightnotice}%
  \def\@evenfoot{}%
}
\def\mycopyrightnotice{%
  {\footnotesize Utkucan Balci and Chen Ling contributed equally to this work.\hfill}
  \gdef\mycopyrightnotice{}
}

\@ifundefined{showcaptionsetup}{}{%
 \PassOptionsToPackage{caption=false}{subfig}}

\makeatother
\usepackage[hyphenbreaks]{breakurl}

\begin{document}

\title{A First Look at Zoombombing}

\author{Chen Ling$^1$, Utkucan Balcı$^{2}$, Jeremy Blackburn$^2$, and Gianluca Stringhini$^1$\\[0.5ex]
 $^{1}$Boston University, $^2$Binghamton University\\
\normalsize ccling@bu.edu, ubalci1@binghamton.edu, jblackbu@binghamton.edu, gian@bu.edu\vspace*{-0.3cm}}\date{}

\maketitle

\begin{abstract}
Online meeting tools like Zoom and Google Meet have become central to our professional, educational, and personal lives.
This has opened up new opportunities for large scale harassment.
In particular, a phenomenon known as zoombombing has emerged, in which aggressors join online meetings with the goal of disrupting them and harassing their participants.
In this paper, we conduct the first data-driven analysis of calls for zoombombing attacks on social media.
We identify ten popular online meeting tools and extract posts containing meeting invitations to these platforms on a mainstream social network, Twitter, and on a fringe community known for organizing coordinated attacks against online users, 4chan.
We then perform manual annotation to identify posts that are calling for zoombombing attacks, and apply thematic analysis to develop a codebook to better characterize the discussion surrounding calls for zoombombing. 
During the first seven months of 2020, we identify over 200 calls for zoombombing between Twitter and 4chan, and analyze these calls both quantitatively and qualitatively.
Our findings indicate that the vast majority of calls for zoombombing are not made by attackers stumbling upon meeting invitations or bruteforcing their meeting ID, but rather by insiders who have legitimate access to these meetings, particularly students in high school and college classes.
This has important security implications, because it makes common protections against zoombombing, such as password protection, ineffective.
We also find instances of insiders instructing attackers to adopt the names of legitimate participants in the class to avoid detection, making countermeasures like setting up a waiting room and vetting participants less effective.
Based on these observations, we argue that the only effective defense against zoombombing is creating unique join links for each participant. 
\end{abstract}
\IEEEpeerreviewmaketitle

\section{Introduction}

One of the earliest promises of the Internet was to enable quick, easy, and real-time communications, not just via text, but also audio and video.
While it took some time, there are now numerous online meeting tools like Skype, Zoom, and Google Meet that are used in a variety of contexts, both personal and professional.
In 2020, society has found itself increasingly reliant on these online meeting tools due to the COVID-19 pandemic, with many business meetings, online classes, and even social gatherings moving online.
Unfortunately, the mass adoption of these services has also enabled a new kind of attack where perpetrators join and deliberately disrupt virtual meetings.
This phenomenon has been dubbed \emph{zoombombing}, after one of the most used online meeting platforms~\cite{brown2020notes,zoombombingwiki}.

To mitigate the threat of zoombombing, security practitioners have begun discussing best practices to prevent these attack from happening or limit their effects. 
These include requiring a password to join online meetings, setting up a waiting room and manually vet participants before letting them in, and not sharing meeting links publicly~\cite{fbi,zoomlog_keep}.
While helpful to keep out casual and unmotivated attackers, there is an inherent tension between tightening the security of online meeting rooms and the need for them to be easily accessible to a number of people, especially in the case of large public events~\cite{brown2020notes}.
Most importantly, devising effective security policies requires a good understanding of the capabilities of attackers and of their modus operandi.
To date, however, the research community lacks a good understanding of how zoombombing attacks are called for and how they are carried out.
For example, it remains unclear how attackers obtain meeting links in the first place.
This type of knowledge is crucial because, for example, protecting against attackers proactively bruteforcing the ID of meeting rooms is very different (and calls for different countermeasures) than mitigating attacks called from insiders.

In this paper, we perform the first measurement study of calls for zoombombing attacks on social media.
We first select ten popular online meeting services, spanning a wide range of target users, from businesses to individuals.
We then analyze the security features that these services offer to their users, with a particular focus on the mechanisms that allow them to restrict and control who can join and participate in the meeting.
We next identify posts that contain online meeting information.
We decide to focus on two online services for this purpose, a mainstream social network like Twitter and a fringe Web community like 4chan, which was shown by previous work to be often involved in harassment attacks against online users~\cite{hine2017kek,mariconti2019you}.
Between January and July 2020, we identify 12k tweets and 434 4chan threads discussing online meeting rooms.
We then apply thematic qualitative analysis~\cite{tseng2020tools} to identify posts that are indeed calling for a zoombombing attack, and to further characterize them.
We identify 123 4chan threads discussing such attacks as well as 95 tweets.
We then adopt a mixed methods approach to perform further analysis. 
We first analyze this dataset quantitatively, looking at temporal properties of these posts and applying natural language processing techniques to better understand the topics of discussion.
We then dig deeper into our qualitative analysis results to get a more nuanced view of the characteristics of the zoombombing phenomenon.
Finally, we discuss our findings in view of existing countermeasures, reasoning about their effectiveness.

\noindent In summary, we make the following key findings:

\begin{itemize}
  \item The majority of the calls for zoombombing in our dataset target online lectures (74\% on 4chan and 59\% on Twitter).
    We find evidence of both universities and high schools being targeted.
  \item Most calls for zoombombing come from insiders who have legitimate access to the meetings (70\% on 4chan and 82\% on Twitter).
    This has serious security implications, because it makes passwords ineffective to protect the meeting rooms as attackers can share them with whoever participates in the attack.
    In some cases we find that the insider shares additional information like names of real students in the class, allowing participants to select those names and make it difficult for teachers and moderators to identify intruders.
  \item Almost all calls for zoombombing target meetings happening in real time (93\% on 4chan and 98\% on Twitter), suggesting that these attacks happen in an opportunistic fashion and that zoombombing posts cannot be identified ahead of time, allowing defenders to prepare.
\end{itemize}

\descr{Disclaimer.} 
Due to their nature, zoombombing messages on social media are likely highly offensive.
In this paper we do not censor any content, therefore we warn the reader that some of the quotes included in the following sections are likely to be upsetting and offensive.

\section{Background}
In this section, we first describe the threat model that we assume for this study.
We then describe how we chose the ten meeting services that we study, and describe their features.

\subsection{Threat Model}
\begin{figure}[t]
	\centering
	\includegraphics[width=\columnwidth]{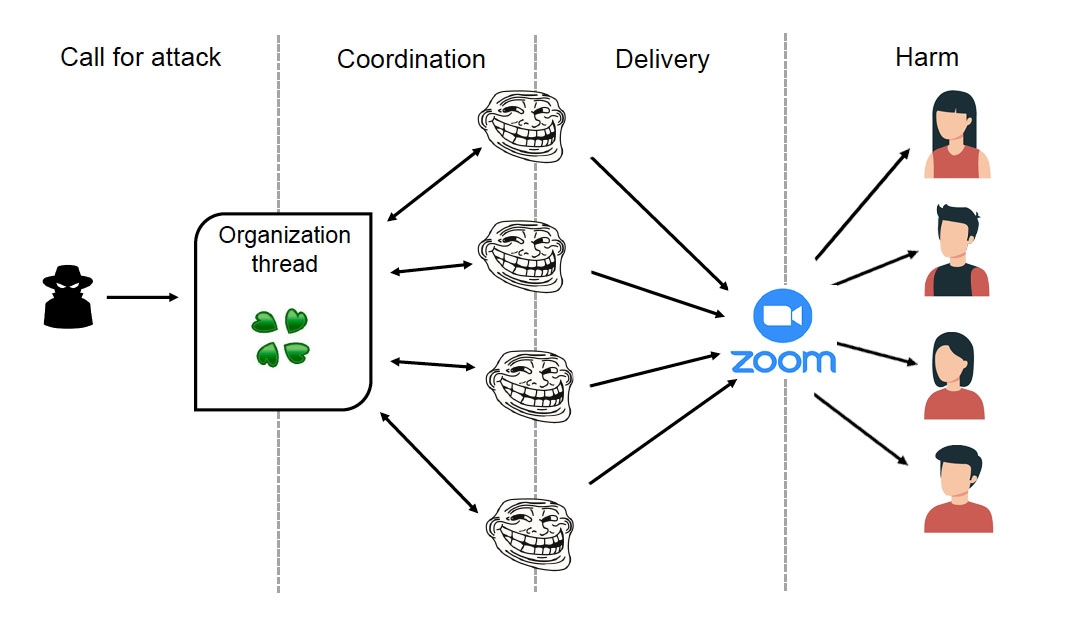}
	\caption{Threat Model for a zoombombing attack. Charlie calls for an attack against a Zoom meeting created by Alice, by creating a thread on an online service (e.g., 4chan). Participants then join the Zoom meeting, report back on the thread about the status of the attack, and harm the legitimate participants to the meeting.}
	\label{fig:model}
\end{figure}
\label{ThreatModel}

We consider a zoombombing attack as being composed of four phases (see Figure~\ref{fig:model}), based on anecdotal evidence of how zoombombing accounts unfold, as well as following empirical evidence reported by previous research that studied coordinated online aggression, trolling, and harassment on other social media platforms (e.g., Reddit, YouTube)~\cite{flores2018mobilizing,hine2017kek,kumar2018community,mclean2019female}. 
Note that in this paper we focus on calls for attacks that aim at attracting multiple participants; single attacks stumbling upon meeting rooms and disrupting them are out of scope.
In the following, we describe the four phases in detail through an example in which Charlie is orchestrating a coordinated attack against a Zoom meeting created by Alice.

\noindent\textbf{i) Call for attack.} 
Charlie obtains information about Alice's Zoom meeting. 
As we will show later, this is often because Charlie is a legitimate participant of the meeting (e.g., a student in an online lecture).
Charlie then posts information about the Zoom meeting on an online service of his choice (starting an \emph{organization thread}), asking other members of the community to participate in a coordinate attack. 
Previous research showed that attacks like this are often organized on polarized Web communities (e.g., /pol/, 4chan's Politically Incorrect Board), where the person calling for an attack posts a link to content on another service that was created by the victim (e.g., a zoom meeting), followed by an invite to the person (e.g., through the phrase ``you know what to do'')~\cite{hine2017kek,mariconti2019you}.

\noindent\textbf{ii) Coordination.} The organization thread created by Charlie now becomes an aggregation point for attackers, who will report additional information and coordinate the attack by replying to the thread.
For example, attackers will post details like a password to access the meeting or personal information about the host. 

\noindent\textbf{iii) Delivery.} The attackers will then join the online meeting and harass the participants, for example sending them hateful messages, shouting profanities, or displaying offensive or indecent images through their webcams \cite{brown2020notes}.

\noindent\textbf{iv) Harm.} The goal of the attack is to cause harm to the group of people. 
Depending on its success and intensity, victims could suffer serious psychological~\cite{fox2017women,hinduja2010bullying} or even physical harm~\cite{macallister2016doxing}.

\subsection{Online Meeting Services}
\label{subsec:meetingservices}

To select a representative set of online meeting tools to study in this paper, we ran Google queries for ``online meeting services'' and manually vetted the results for Web pages that actually advertised a service (excluding, for example, news articles talking about a certain meeting platform).
After this process, we obtained the list of the ten highest ranked meeting tools.
These services are \href{https://zoom.us}{Zoom}, \href{https://hangouts.google.com}{Hangouts}, \href{https://meet.google.com}{Google Meet}, \href{https://www.skype.com/en/}{Skype}, \href{https://jitsi.org}{Jitsi}, \href{https://www.gotomeeting.com}{GotoMeeting}, \href{https://www.microsoft.com/en-us/microsoft-365/microsoft-teams/group-chat-software?OCID=AID2100233_SEM_X0L4TQAABfN-VXsE:20200823181421:s&msclkid=921df13fc06c168126acae7d382755af&ef_id=X0L4TQAABfN-VXsE:20200823181421:s}{Microsoft Teams}, \href{https://www.webex.com}{Cisco Webex}, \href{https://www.bluejeans.com}{Bluejeans}, and \href{https://starleaf.com}{Starleaf}.

In the following, we describe the general characteristics of each of these services (see Table~\ref{tab:background}).
We then analyze the security relevant features offered by the various platforms (e.g., whether they allow hosts to set a password for meetings).
We are particularly interested in understanding what characteristics of a service might make it a popular target platform for attackers, or might reduce the risk for a successful attack.

\begin{table*}[t]
  \begin{scriptsize}
    \begin{center}
	\begin{tabular}{lllllll}
		\toprule
		Platform & Est. & Headquarters & Parent Company & Target Users & User base & Plan \\
		\midrule
		 Zoom & 2011 & US & - & Both individual and business & 300M & Free, upgrade available starts from \$15/month\\
		 Meet & 2017 & US & Google & Both individual and business & 100M & Free, upgrade available starts from \$12/month \\
		 Webex & 1993 & US & Cisco & Business & 324M & Free, upgrade available starts from \$13.5 /month\\
		 Jitsi & 2017 & AU & Atlassian & Both individual and business& - & Free\\
		 Skype & 2003 & US & Microsoft & Both individual and business & 100M & Free, charge for phone calls\\
		 GotoMeeting & 2004 & US & LogMeIn & Business & - & Starts from \$12/Month\\
		 Teams & 2017 & US & Microsoft & Business & 75M & Free, upgrade available starts from \$5 per user/month\\
		 Hangouts & 2013 & US & Google & Individual & 14M & Free, charge for phone calls\\
         Bluejeans & 2009 & US & Verizon & Business & - & Starts from \$12/Month\\
		 Starleaf & 2008 & UK & - & Business & 3,000 & Free, upgrade available starts from \$14.99 /month\\
		\bottomrule
	\end{tabular}
    \end{center}
  \end{scriptsize}
	\caption{Overview of the ten online meeting services studied in this paper.}
	\label{tab:background}
\end{table*}

\noindent\textbf{Length of operation.} Half of our ten services were established after 2010, with the notable exception of Webex which started in the 90s.
Major tech companies like Microsoft, Google, and Cisco have their own solution, with Microsoft and Google having two of them (Skype and Teams for Microsoft and Hangouts and Meet for Google).
While Google started retiring Hangouts in October 2019, we will later show that this platform is still very much used and many meeting links to it are posted on social media.
There are also companies that focus on online communication services, like Zoom and Starleaf.
During the coronavirus pandemic, when millions of people have been forced to work, learn, and socialize remotely, Zoom has risen to the top, with over 300 million daily participants in virtual meetings, and also becoming the top target of attack; hence the phrase ``zoombombing.''

\noindent\textbf{User base.} 
Most of the online meeting services are aimed at business users.
While Hangouts is the only service specifically devoted to individuals, five of them are geared towards both business and individual users.
Based on the most current data~\cite{ciscoaddmoreuser,googlemeetaddmoreuser,skypeaddmoreuser,zoomaddmoreuser} (July 2020), four of our selected online meeting services have a user base of over 100M (Zoom, Meet, Skype, and Webex).
We hypothesize that the user base of a service plays a role in which services get attacked the most.

\noindent\textbf{User plan.} 
Most online meeting services provide free accounts for individuals and small companies.
GotoMeeting and Bluejeans, however, exclusively target business consumers (charging hosts \$12/month) and do not provide free accounts.
Teams paid plans are somewhat different, as they are based not on a per-host basis, but on a per-user basis.
Google Hangouts and Skype are free, but charge for phone calls to local numbers.

\begin{table*}[t]
	\centering
	\begin{scriptsize}
	\begin{tabular}{lllllllc}
		\toprule
		Platform & Requires account to join in & Max particp. & Max time & Allows password & Allows waiting room & one-time unique link & Mute upon entry\\
		\midrule
		 Zoom & Yes & 100 & 40min & Yes & Yes & for each particp. & Yes\\
		 Google Meet & Yes & 100 & Unlimited & No & Yes & No & No\\
		 Webex & Yes & 100 & 50min & Yes & No & for each particp. & Yes\\
		 Jitsi & No & 75 & Unlimited & No & No & Yes & No\\
		 Skype & Yes & 50 & Unlimited & No & No & No & No\\
		 GotoMeeting* & No & 26 & Unlimited & Yes & Yes & Yes & Yes\\
		 Teams & Yes & 4 & Unlimited & No & No & Yes & No\\
		 Hangouts & Yes & 25 & Unlimited & No & No & No & No\\
         Bluejeans* & Yes & 50 & Unlimited & Yes & Yes & Yes & Yes\\
		 Starleaf & Yes & 20 & 45min & No & No & Yes & Yes\\
		\bottomrule
	\end{tabular}
	\end{scriptsize}
	\caption{Comparison of the features offered by the online meeting services studied in this paper to free accounts. Services marked with * do not provide a free version and are only available to hosts who pay a subscription.} 
	\label{tab:capability}
\end{table*}

\noindent\textbf{Features.} 
We next analyze the features that are specific to each online meeting platform, with a particular focus on the security measures that they put in place to prevent zoombombing. 
To this end, we compare the features offered to free accounts.
Since GotoMeeting and Bluejeans do not provide free accounts, they are excluded from this comparison, since we could not create meetings to check their capabilities. 
An overview of the features offered by each platform is reported in Table~\ref{tab:capability}.

First, we look at the security features offered by the meeting platforms.
Nine of the ten services require an account to join a meeting. 
This is done to prevent attackers from flooding meeting rooms and provide some accountability, e.g., suspending misbehaving accounts.
Only Jitsi does not require a registration to join meetings.
Authentication-wise, the security model of online meeting services is the following: anyone with an account on the platform and who knows the meeting ID can join the meeting.
This is not dissimilar to other security sensitive services that have been studied by the community in the past, from online document editing~\cite{kaleli2019perils} to file download platforms~\cite{lauinger2013holiday}.
To prevent anyone knowing the meeting ID from joining a room, Zoom, Webex, GotoMeeting, and Bluejeans allow hosts to specify a password participants need to provide upon joining. 
Only Zoom and Google Meet allow a waiting room for hostswhi to check identity of participants.
Google Meet automatically admit participants whose accounts were included in the invitation list into the meeting room and puts others in a waiting room, allowing the host to let them in manually.
Only Zoom and Webex provide a registration system with one-time unique links per registrant, which can help restrict and trace participants.
Generally, other meeting services use unique links for each meeting, with Google Hangouts and Google Meet allowing a link to be reused within a 90 day period.
Skype does not have a one time unique link function.
Due to privacy concerns, Google Meet, Google Hangouts, and Jitsi do not allow host to mute all participates~\cite{googlemuteall,Jitisimuteall}.
Google Meet only allows educational accounts to mute participants~\cite{googleeducationmute}.

Second, we look at whether services limit the number of users that can join a meeting, as well as the maximum duration of a meeting for free users.
All the services under study have a participant limit in their free version.
Zoom, Google Meet, and Webex limit meetings to 100 participants, and Teams only supports four attendees in its free version.
When looking at the maximum duration of a meeting, we find that three services (Zoom, Webex, and Starleaf) limit meetings to between 40 and 50 minutes for free users.

\section{Datasets}
\label{sec:datasets}

In this section, we describe the datasets that we used in this paper as well as our data collection process.
We first discuss how we identify social media posts containing links to meeting rooms.
We then discuss the online services that we collect data from.

\noindent\textbf{Identifying posts containing meeting URLs.}
To identify posts that contain meeting URLs on the online services that we monitor, we first identify the DNS domains that are used by the platforms that we are studying.
To avoid simple attempts to evasion, we used regular expressions that only considered alphanumeric characters and dots. 
In the case of Zoom meetings not shared by URL but instead via meeting ID, after lowercasing and removing non-alphanumeric characters in the posts, we searched for a pattern with `id' followed by at least nine consecutive digits by using regular expressions.
We then further filter these by only including posts with the keyword `zoom' in them.
\noindent\textbf{4chan.}
4chan~\cite{nagle2017kill} is an imageboard where users can start a thread anonymously, with other users commenting on it. 
4chan is organized in boards that either cover different topics of discussion (e.g., Anime \& Manga, Sports) or are created to host more generic discussion (e.g., Politically Incorrect, Random).
Unlike traditional online services, threads on some of the 4chan boards are \emph{ephemeral}, and only a fixed number of threads is alive at a time.
Once a new thread i created, the active thread that has least recently been used is removed from the catalog of live threads.
Previous research showed that 4chan is a popular platform used by miscreants to carry out abuse, such as organizing coordinated harassed campaigns~\cite{hine2017kek,mariconti2019you,nagle2017kill}.
We therefore hypothesize that zoombombing is widespread on the platform.

We developed a custom crawler following the same methodology of previous research on 4chan~\cite{hine2017kek,papasavva2020raiders}, and collected all posts between January 1st, 2020, and July 24th, 2020.
We then identify posts containing online meeting links and invitations following the methodology discussed in the previous section. 
Every time we identify a post containing information about a meeting, we pull the entire thread.
In total, we identify 47,221 posts from 434 threads with a URL or an ID for at least one meeting platform room.

\noindent\textbf{Twitter.}
Twitter~\cite{kwak2010twitter} is a microblogging social media platform on which registered users can share posts publicly or privately. 
While private accounts can only reach their followers, public accounts can reach any user on Twitter. 
The posts are called ``tweets'' and can be re-shared (retweeted) by other users to share with their followers. 
Tweets can contain ``hashtags'' where users can put the ``\#'' symbol at the beginning of a word. 
By using the same hashtags, people can create trends, which can also be used to look up tweets on the same topic.

Leveraging the Twitter streaming API, a public service that makes a random 1\% sample of all tweets posted worldwide, we identified 12,077 tweets containing links or IDs to online meeting rooms. 
These tweets were posted between January 1st, 2020, and July 18th, 2020.
Note that due to limitations in the Twitter API we could not retrieve any replies to tweets containing meeting IDs.

\noindent\textbf{Ethics.}
We acknowledge that data from social media can contain personal information.
We adopted standard best practices to ensure that our study followed ethical principles~\cite{bailey2012menlo,rivers2014ethical}
In particular, we did not try to further de-anonymize any user. 
Since this work only involved publicly available data and did not require interactions with participants, it is not considered human subjects research by our institution. 

\section{Identifying Zoombombing Threads}
\label{section:annotation}
While it is relatively straight forward to automatically find posts that include links to meetings, the challenge is in determining the intent behind the link being posted, and in particular whether the post is calling for a zoombombing attack.
We expect that most meeting links on social media are posted with benign reasons; therefore, to carry out this study we need a way to separate harmless posts from those that are calls for zoombombing.
Since zoombombing is a human driven phenomenon, developing automated techniques to identify posts calling for attacks is challenging and prone to false positives and false negatives.
To avoid these issues, we perform manual annotation of all posts in our dataset, with the goal of identifying a reliable ground truth dataset.

In this section, we develop a codebook to guide the thematic annotation process for our 4chan and Twitter datasets.
We break the development of this codebook in two phases.
First, we perform a binary labeling to determine if posts are indeed calls for zoombombing or not.
As a second step, we further characterize the posts and threads that contain zoombombing invitations, with the goal of understanding the behavior of attackers and the targets that they choose.

To build our codebook and perform annotation we follow the same methodology described in recent security research~\cite{tseng2020tools}, in which the authors studied posts from online infidelity forums and their relation with intimate partner surveillance tools and tactics. 
More precisely, we follow these four steps:
\begin{enumerate}
  \item Four researchers independently screened our dataset and produced initial codes using thematic coding~\cite{braun2006using}.
\item We then discussed these initial codes and went through multiple iterations, using a portion of the data to build a final codebook. 
The process continued until the codebook reached stability and additional iterations would not refine it further.
\item To investigate the common agreement on the codebook by multiple annotators, we have them rate a portion of our dataset and discuss disagreements until a good agreement is reached.
\item We split the rest of our dataset and each annotator labels one portion of it.
\end{enumerate}

\noindent We next describe our process and our codebook in more detail.

\subsection*{Phase I: labeling zoombombing content}

As we mentioned, the first phase of our annotation process deals with identifying social media posts and threads that contain an invitation to zoombombing.
We start by labeling 4chan threads.
Following the methodology from~\cite{tseng2020tools}, we first randomly choose 10 threads from the 470 threads that contain a link to a meeting room, and have each author of the paper review them and discuss them together to build a shared understanding of what a zoombombing invitation looks like.
From this initial dataset, the authors agreed that two threads were ``bombing'' threads (i.e., they were encouraging/calling for a zoombombing) while the remaining eight were not (i.e., ``non-bombing'').

We then aim to test each author's ability to independently identify bombing threads.
To this end, we chose 20 additional threads (balanced as per the overall distribution of meeting platform links on 4chan), and had each author label them as either bombing or non-bombing.
We used the following definition to make a decision: \emph{a zoombombing thread should include an invitation to bomb along with a URL to a meeting room or a meeting ID}.
One interesting caveat here is that while discussing the initial set of threads we noticed that the invitation to bomb did not necessarily appear in the same post as the meeting link itself, and thus we added the following additional condition where applicable: \emph{the same user posted the link or meeting ID and the textual invitation to bomb, even if they were not in the same post}. Note that although users on 4chan are anonymous, users are giving a unique ID that identifies them within the same thread~\cite{hine2017kek}.
It is important to note that invitations to bomb are not necessarily made in an overt fashion.
4chan's users are well known to use coded language and slang~\cite{mariconti2019you}, and thus we relied on our domain expertise when coding posts that include phrases like ``you know what to do'' and ``do ya thing.''
Finally, because of the overall uncertainty of things, we decide to be conservative and label any threads we are unsure about as non-bombing.
A typical bombing invitation looks as follows:

\begin{mdframed}[style=MyFrame,nobreak=true,align=center]
\begin{quote}
``[ZOOMURL] My English class, come in and trolley for a while.''
\end{quote}
\end{mdframed}

Four authors of this paper independently coded each thread to determine whether it is bombing related or not.
From this testing phase of 20 threads, we calculated the Fleiss' agreement score between the annotators and found perfect agreement ($\kappa = 1.0$)~\cite{fleiss1971measuring,fleiss2013statistical}.
This indicates that all authors were able to reliably identify zoombombing threads.
From here, we expand our annotation to the full dataset of 434 threads, split evenly between the four annotators.

\begin{figure}
	\centering
	\includegraphics[width=0.7\linewidth]{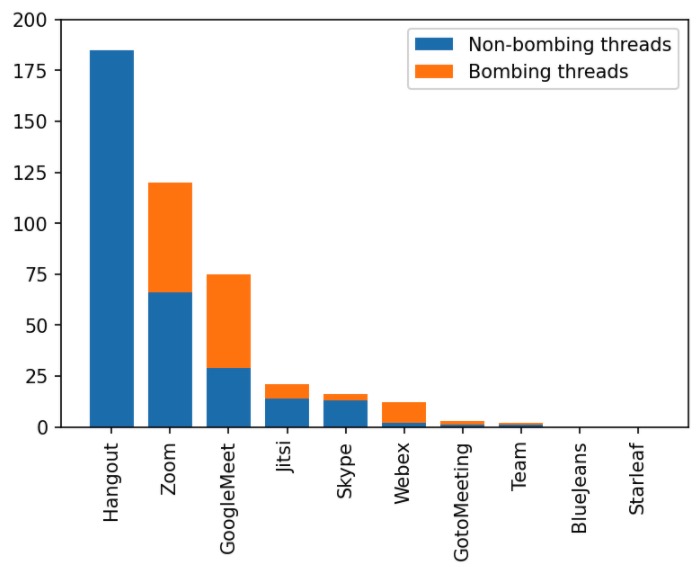}
	\caption{Ratio of bombing and non-bombing posts on 4chan.}
	\label{fig:4chanbombing}
\end{figure}

In the end, we find that 123 of the 434 threads in our 4chan dataset are bombing threads.
As seen in Figure~\ref{fig:4chanbombing}, nearly half (43.96\%) of the Zoom meeting links in our dataset were determined to belong to a bombing thread, and a majority (59.72\%) of Google Meet links appeared in bombing threads.
On the other hand, Google Hangouts and Skype links are mostly posted with benign intentions.

\begin{figure}
	\centering
	\includegraphics[width=0.7\linewidth]{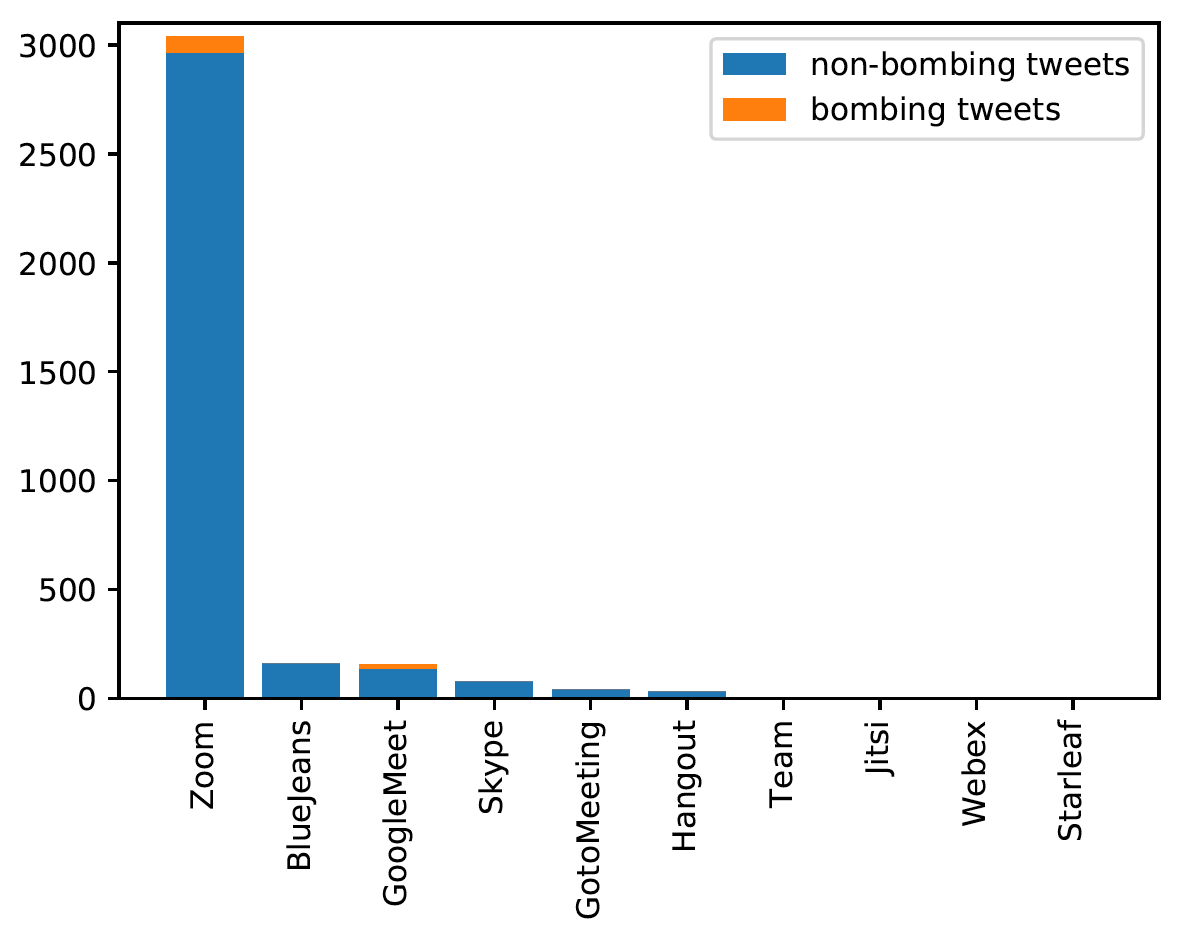}
	\caption{Ratio of bombing and non-bombing tweets on Twitter.}
	\label{fig:twitterbombing}
\end{figure}

We follow the same labeling procedure for Twitter.
From our preliminary screening of the tweets, we find that a large portion are non-English.
Thus, we restrict our analysis to English tweets only from our total 12,077 tweets, which leaves us with 3,510 candidate tweets.

A challenge that we face when labeling tweets is that Twitter is a much different platform than 4chan in its user base and general tone. 
4chan is dominated by trolling and irony, and veiled calls to join meetings can often be interpreted as bombing invitations.
Here is an example of a bombing invitation from 4chan:

\begin{mdframed}[style=MyFrame,nobreak=true,align=center]

\begin{quote}
``Ok retards, this is an id of a zoom web lessons. Do your worst [ZOOM ID] [ZOOM PASSWORD].''
\end{quote}
\end{mdframed}

On the other hand, Twitter is a general audience social network, therefore we expect most meeting invitations to be benign.
For example, this is a bombing invitation from Twitter:

\begin{mdframed}[style=MyFrame,nobreak=true,align=center]
\begin{quote}
``Raid this class as fast as u can....\\
\#zoomcodes \#zoomclasscodes \#zoomclass \#zoom [ZOOMURL]''
\end{quote}
\end{mdframed}

To reflect this difference and avoid potential false positives, we decide to be stricter when determining if a tweet is a zoombombing invitation.
More precisely, a bombing tweet needs to meet the following two criteria:
\begin{itemize}
    \item An invitation to bombing with a link (invitation text usually comes with a link)
    \item A clear indication of bombing, such as ``raid,'' ``bomb,'' ``troll,'' ``discord,'' ``disruptive,'' and ``make fun of it.''
\end{itemize}
As with 4chan, we are generally conservative in our labeling and default to non-bombing in uncertain cases.

From the 3.5K English tweets, we randomly sample 500 so all services were equally represented (i.e., balanced with respect to services).
From this 500, we manually select 20 tweets, which four coders independently determined whether they were a bombing tweet or not.
The inter-rater reliability again shows perfect agreement (Fleiss' $\kappa = 1.0$).
Because of the high agreement scores on the initial testing set, as well as the agreement on the 4chan ratings, we had a single annotator label the remaining 3,490 tweets in this dataset.
Note that this is a much quicker process than on 4chan, since the coder had to look at single tweets instead of entire, and often long, threads.

In the end, we find that 95 out of the 3,510 candidate English tweets are bombing tweets.
From Figure~\ref{fig:twitterbombing} we can see that zoombombing on Twitter is less pervasive than on 4chan.
In particular, of the 3,039 Zoom related candidate tweets, 75 are labeled as bombing, and 20 of the 157 Google Meet tweets are bombing.
We found no bombing tweets for the other eight meeting tools.

\subsection*{Phase II: Characterizing zoombombing}
While labeling threads and tweets as bombing or not is vital to understanding the problem, it does little to characterize the actual bombing activity itself.
In this phase we aim to understand the \emph{process} of a bombing event by analyzing the behavior that goes on in bombing threads.

We began by having four annotators go through the labeled bombing threads/tweets as determined by the Phase I labeling.
This was a relatively loose process where the goal was to get a general sense of what is going on.
Next, the annotators met and discussed their observations.
In general there was agreement between the annotators of a clear trend of insider complicity in bombing of online classes in particular.
After several rounds of discussion, we derived four, high level properties relevant to zoom bombing threads and tweets: 1)~thread structure (only applicable to 4chan threads), 2)~link information, 3)~invitation information, and 4)~interaction (only applicable to 4chan threads).

\descr{Thread structure:}
New threads on 4chan are created when a so called ``Original Poster'' creates an ``Original Post'' and the thread constitutes replies to this post (\textbf{NB:} 4chan threads are \emph{flat})~\cite{hine2017kek}.
Thus, the first post in a thread usually represents the topic of the thread.

We code the following characteristics of a thread:
\begin{enumerate}
    \item Whether the content of the first post is a zoombombing invitation. This indicates whether or not the thread was created primarily to act as a bombing thread as opposed to organically evolving into one.
    \item The length of the thread (i.e., the number of posts), which indicates the thread's popularity.
    \item The number of bombing invitation links, which is indicative of how the thread evolved with respect to bombing.
\end{enumerate}

\descr{Link information:}
According to our definition of a bombing thread/tweet, both 4chan and Twitter posts need to include a video conference invitation link or meeting ID to be considered a bombing thread.
For certain meeting platforms (e.g., Zoom) we can derive two additional pieces of information from meeting links directly: 1)~\emph{institutional information} (i.e., who is hosting the meeting) and 2)~\emph{password protection}.

For some platforms, we can automatically identify password-protected links by looking at a password parameter in the URL (e.g., \url{https://zoom.us/j/123456789?pwd=12345aAbBcC678}).
When coding messages manually, we also look at the presence of passwords in the text of posts. 
Institutional information provides us additional information on the victims of attacks.
To gather this information, we need to manually look at the URL (e.g., \url{http://UNIVERSITY.zoom.us/j/XXXXXX}, and search for its associated institution.
We record each institution, its type (e.g., University), and country.

\descr{Invitation information:}
As noted previously, there are plenty of legitimate reasons to post a link to a video conference, and thus a posted link itself is not sufficient to say that an attack has occurred; this is why we require additional text calling for an attack.
During our initial examining, we noticed that there was often additional information embedded in the bombing invitation itself, e.g., temporal details as well as hints at the existence of insiders.

\begin{mdframed}[style=MyFrame,nobreak=true,align=center]
\begin{quote}
``[ZOOMURL] this class is up the tuesdays at 11:00 am UTC-5 crash this class plz.''
\end{quote}
\end{mdframed}

For temporal information, we manually read the bombing invitation and label the meeting time according to three codes 1)~\emph{future event}, where the poster indicates the attached link will be active at some point in the future, 2)~\emph{live event}, where the poster indicates the meeting link is active and that bombers should join ``now,'' and 3)~\emph{not sure}, where there was no clear indication of when the link would be active.
This temporal information is an indicator as to whether or not a bombing attack has been planned, or if it is an opportunistic attack.

Our preliminary analysis indicated that many zoombombing invitations are created by insiders, for example students in the case of college classes.
To better understand insider complicity, we label each bombing post or thread as either 1)~\emph{insider} or 2)~\emph{non-insider}.
To be labeled as \emph{insider}, the bombing invitation should include text like ``my teacher'' or ``our class,'' provide a password for the video conference (either explicitly in post text or implicitly in the link to the meeting), or give suggestions on what names bombers should select when joining the call (a tactic used to make it harder for legitimate meeting attendees/hosts to determine that joining bombers are not supposed to be there).
Annotators recorded the details of what led to any \emph{insider} label applied.
Again, we conservatively label threads as \emph{non-insider} if there is any doubt.

\descr{Interaction:}
For 4chan, we are able to collect entire threads discussing zoombombing.
For these threads, we read the whole thread and record the following characteristics of the thread discussion:
\begin{itemize}
  \item Time interval: the interval between the bombing invitation post and the first interaction post by other users (this characteristic is automatically calculated);
    \item Problem feedback: participants reporting problems about their zoombombing attempts, for example being unable to join the meeting room, or being kicked out by the host;
    \item Toxic speech: participants insulting the host of the meeting with profanities or hate speech;
    \item Crime scene feedback: reports on successful attacks with details on what happens;
\end{itemize}

For phase II, four raters independently rated 20 randomly chosen threads from 123 bombing 4chan threads and 20 random tweets from 95 bombing tweets from Twitter.
Inter-rater reliability showed a perfect agreement in both sets of threads (Fleiss' Kappa 1.0).
We then split the rest of the dataset into 4 groups, with each rater separately coding one group.

\section{Quantitative Analysis}

To better understand the zoombombing phenomenon, we first start by quantitatively analyzing the 123 4chan threads and 95 tweets that we identified as part of the coding process, comparing them with posts and threads containing non-bombing meeting links.
We focus our analysis on three aspects: 1)~understanding which services are targeted the most by zoombombing 2)~examining how zoombombing unfolds temporally and 3)~using natural language processing techniques to quantify the content of zoombombing threads.

\subsection{Targeted services}

We observe that the platforms with a larger user base (see Table~\ref{tab:background}) seem to be attracting more zoombombing attacks.
In particular, we find 129 bombing links on Zoom, 66 on Google Meet, 10 on Webex, 7 on Jitsi, 3 on Skype, 2 on GoToMeeting, and 1 on Teams, while there are none for Hangouts, Bluejeans, and Starleaf.

\subsection{Temporal Analysis}

\begin{figure}
	\centering
	\includegraphics[width=0.8\columnwidth]{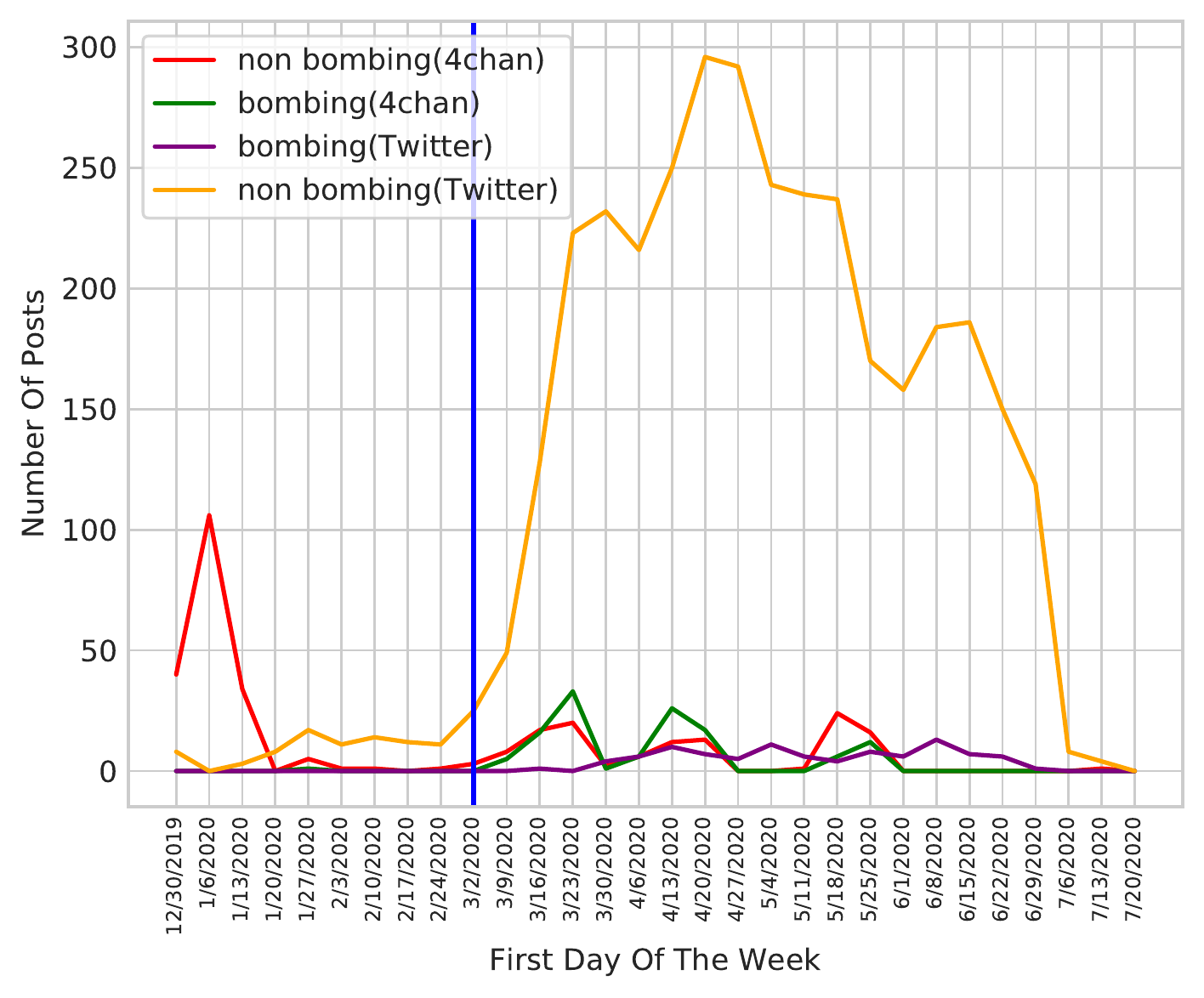}
	\caption{Number of posts per week for bombing \& non-bombing threads and tweets. The vertical line indicates the beginning of the COVID-19 lockdown in the United States (on the week of 3/2/2020, when several West Coast US universities started going online.)}
	\label{fig:no_of_posts_week}
\end{figure}

Figure~\ref{fig:no_of_posts_week} plots the weekly occurrences of bombing and non-bombing posts on Twitter and 4chan.
From the figure, we see that posts with meeting links became more prevalent (especially on Twitter) as the COVID-19 shutdown began in March 2020 (shown in the figure with blue line\footnote{https://www.insidehighered.com/news/2020/03/09/colleges-move-classes-online-coronavirus-infects-more}).
On 4chan, we observe a spike in benign posts containing meeting links around New Years Eve 2020, attributable to users organizing social gatherings as well as increased activity of a far-right group on the following week. 
Generally speaking, zoombombing as a phenomenon barely existed before the quarantine.
We observe a decline of the phenomenon in June 2020, potentially linked to school holidays; this is in line with the fact that we observe that most calls for zoombombing target school lectures and college classes, as discussed later in Section~\ref{subsec:quality1}.

\begin{figure}
	\centering
	\includegraphics[width=0.8\columnwidth]{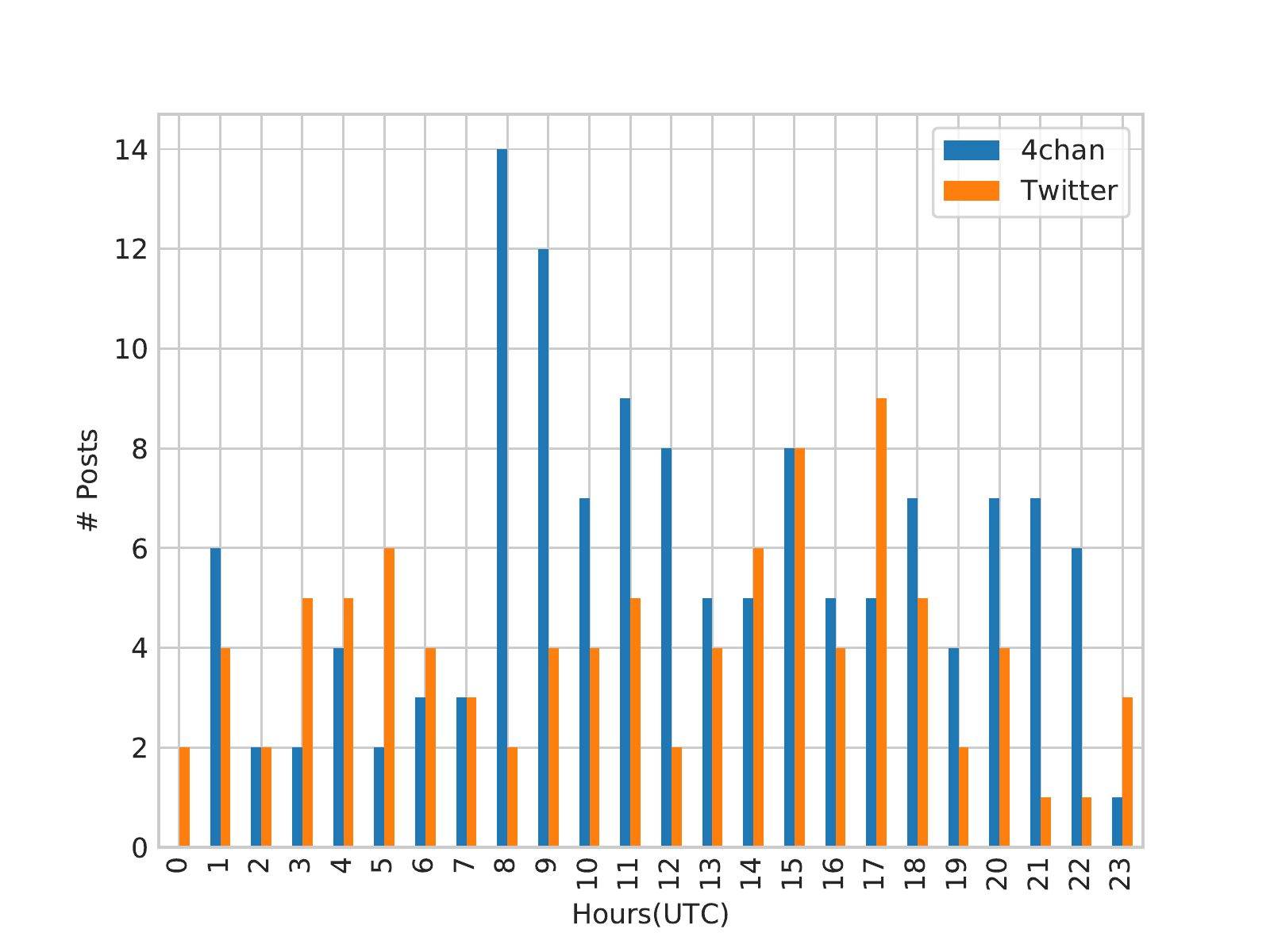}
	\caption{Hour Distribution of zoombombing posts. Note that we did not discard multiple posts that contain the same zoombombing link.}
	\label{fig:posthour}
\end{figure}

Next, we plot the number of posts per hour of the day for 4chan posts and tweets with bombing links in Figure~\ref{fig:posthour}.
On Twitter, we find that zoombombing activity does not exhibit clear diurnal patterns. 
On 4chan, bombing posts are mostly shared from 08:00 to 23:00 UTC. 
We did not encounter any zoombombing tweet that specified a location and only 13 zoombombing posts had country information on 4chan (8 USA, 1 Indonesia, 1 Bulgaria, 1 Turkey, 1 Chile and 1 Italy). 
Considering the lack of diurnal patterns in Figure~\ref{fig:posthour}, we can derive that zoombombing calls are not a localized problem. 

\descr{Temporal analysis of 4chan threads.}
To better understand zoombombing behavior, we analyze the threads on 4chan where zoombombing links were posted. 
This allows us to get a quantitative understanding of how discussion of zoombombing activity unfolds on the platform.
Based on our manually labeled dataset, we extract 123 threads, which contain 2,693 total posts.
We compare these 123 threads to the 311 threads (44,528 posts) that included a meeting link but were \emph{not} bombing threads.
Finally, we also compare to a baseline of 4chan posts chosen by sampling threads at random (without replacement) on a per-day basis such that we have the same number of baseline threads per day as we have threads where a meeting link was posted.

\begin{figure}
	\centering
	\includegraphics[width=0.8\columnwidth]{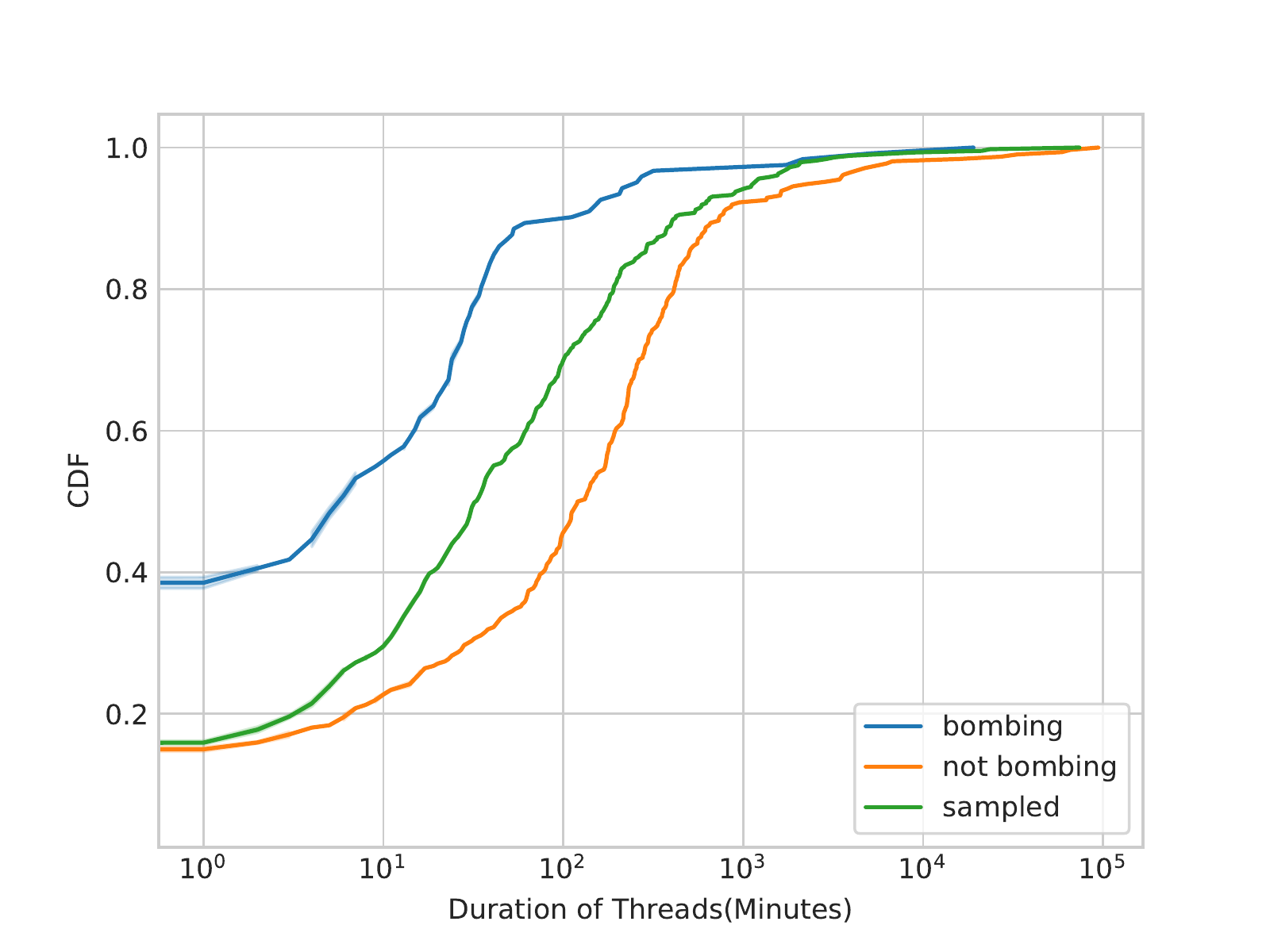}
	\caption{Duration of threads on 4chan.}
	\label{fig:duration}
\end{figure}

Figure~\ref{fig:duration} plots the cumulative distribution function (CDF) of the duration of threads in our dataset (defined as the difference in the timestamp of the last post and the timestamp of the original post).
Recall that threads on 4chan are ephemeral, and once a thread is not active for a while it gets pruned and no further posts can be made~\cite{hine2017kek}.
From the figure, we observe that bombing threads have a shorter lifetime than other threads: 50\% of bombing threads are active for less than 5 minutes, compared to 30 minutes for randomly sampled threads, and two hours for non-bombing threads.
That said, we do have a long tail with about 10\% of bombing threads lasting over 2 hours, compared to 7 hours for sampled threads and 12 hours for non-bombing threads.

\begin{figure}
	\centering
	\includegraphics[width=0.8\columnwidth]{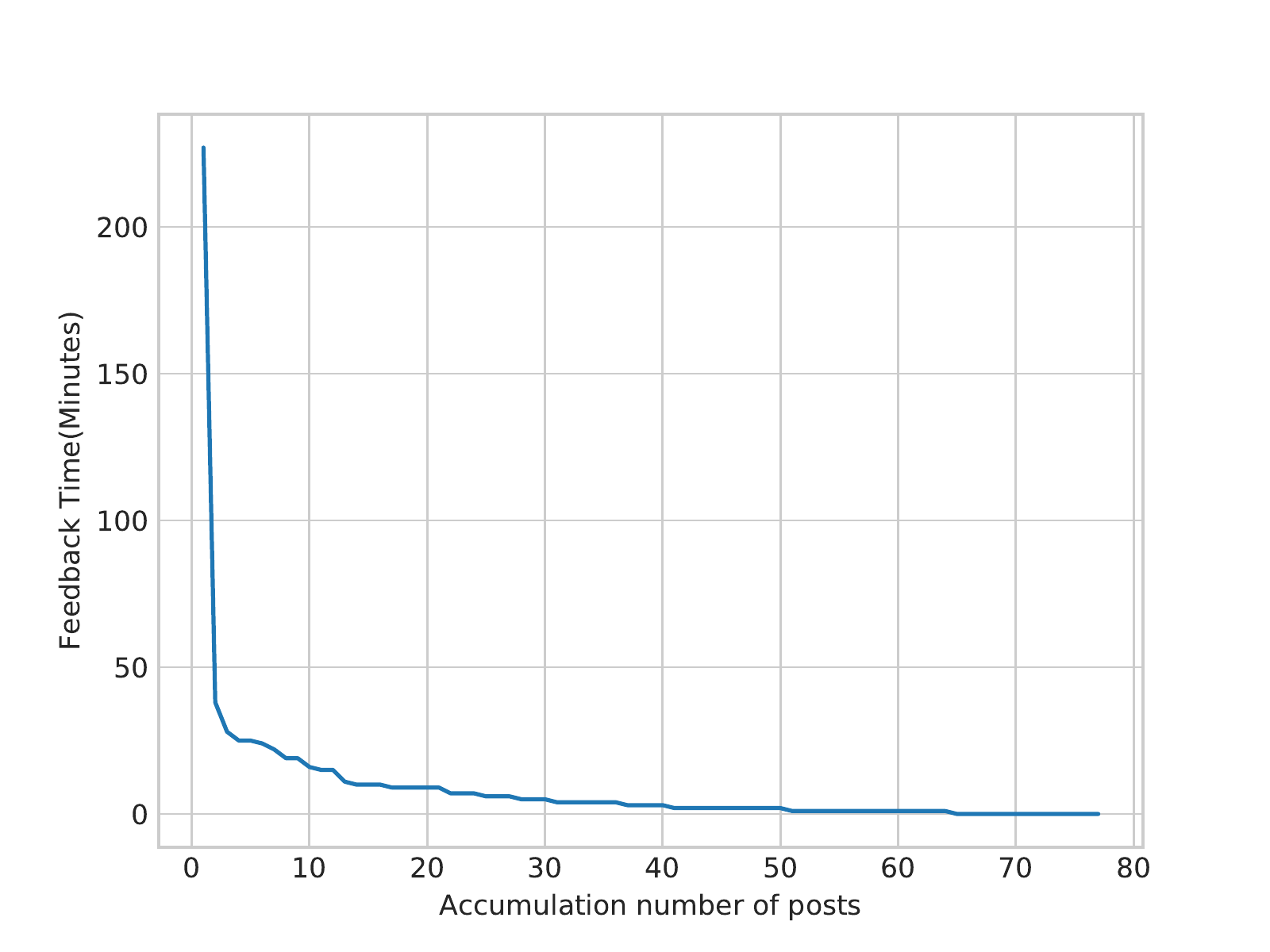}
	\caption{Feedback time between the posting of a zoombombing invitation on 4chan and the first reply to the thread.} 
	\label{fig:feedbacktime}
\end{figure}

\begin{figure}
	\centering
	\includegraphics[width=0.8\columnwidth]{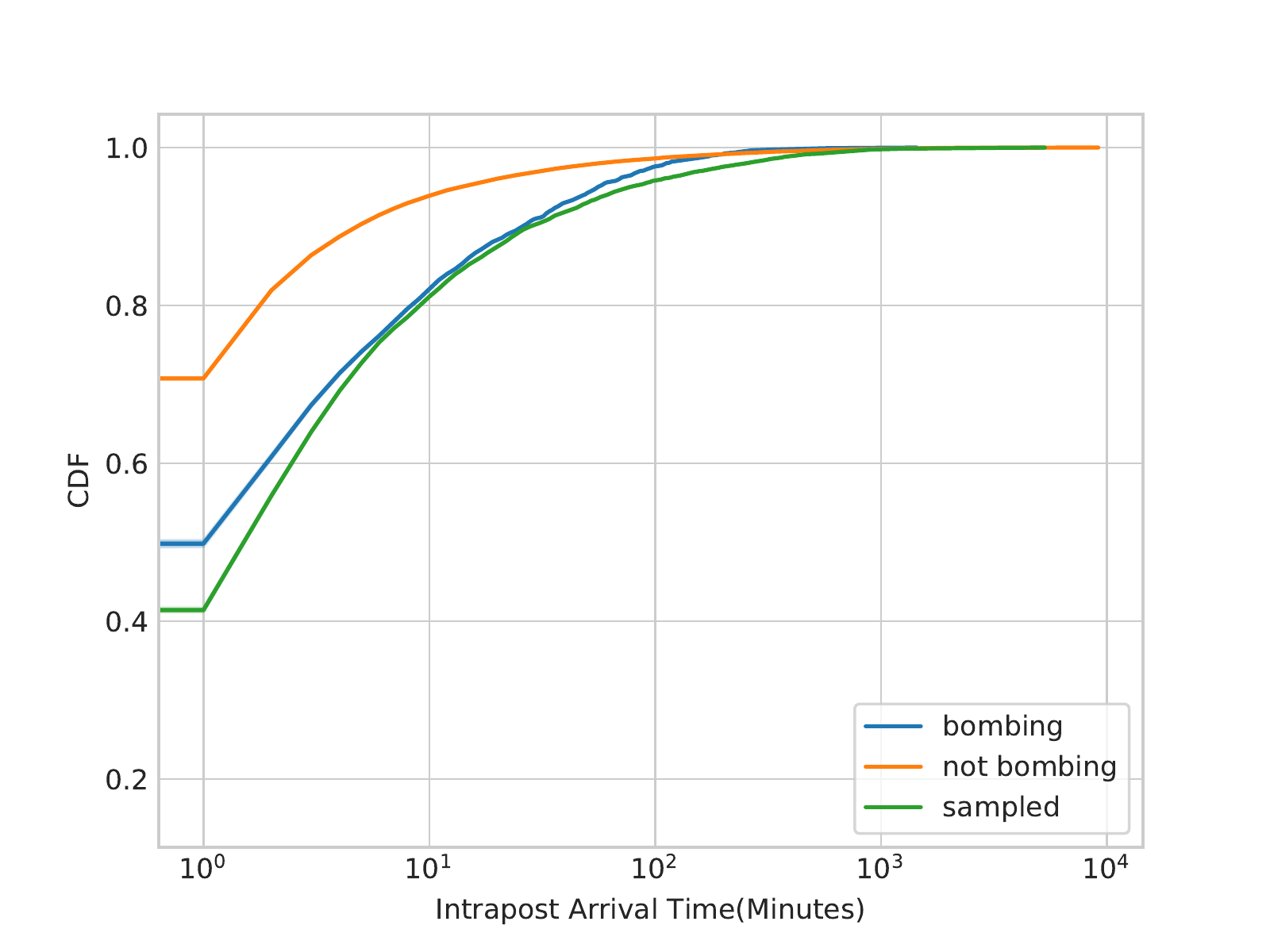}
	\caption{CDF of Interpost Arrival Times for bombing \& non-bombing threads}
	\label{fig:intraparival}
\end{figure}

In our threat model, threads become an aggregation point for attackers, and so understanding the feedback Charlie receives from the bombers he is trying to recruit is important.
Thus, Figure~\ref{fig:feedbacktime} plots the delay between the bombing link being posted on 4chan and the first reply.
From the figure, we see that 79\% of zoombombing threads receive their first reply within 10 minutes.
One explanation for this is that calls for zoombombings might be time sensitive; indeed in Section\ref{subsec:quality2} we show that many of our attackers are inviting bombers to join live meetings/classes.
We then look at the interpost arrival time between each post in a thread.
Similarly, Figure~\ref{fig:intraparival} plots the CDF of interpost arrival times, which is the time between consecutive posts in threads, for bombing and non-bombing threads.
For most threads the elapsed time between consecutive posts in bombing threads is similar to sampled threads while being higher compared non-bombing threads.
One explanation for this is that non-bombing meeting links tend to be posted to organize social gatherings, and thus tend to show up in more popular, faster moving threads.
An alternative explanation is that while the zoombombing attack is happening 4chan users are slower in replying in the thread because they are busy performing malicious activities in the meeting room. 

\subsection{Characteristics of zoombombing links}
\label{subsec:characterzblink}

In this section we focus on what we can learn by analyzing the zoombombing links, in particular whether they contain information about the victim organizations and if they include a password as a URL parameter.

\noindent\textbf{Targeted organizations.}
We want to understand what organizations are victims of zoombombing.
Two of the services (Zoom and Webex) that we study allow organizations to set up a subdomain that identifies them (for example \url{https://virginia.zoom.us/j/123456789} to identify the University of Virginia on Zoom and \url{https://pacificbuddhistacademy.my.webex.com} for the Pacific Buddhist Academy).
We find that most of the zoombombing links posted on 4chan and Twitter are generic and do not contain subdomains that are specific to any organization: only 12 links contain specific subdomains to 10 institutions, and 2 links contain specific subdomains to 1 institution on Twitter.
In particular, we find that 8 zoombombing links on 4chan belong to education institutions while there are none on Twitter.
One of these is a high school located in the US (Evergreen PS in Washington), four are universities in the US (e.g., Arizona State University), and three are universities outside the US (e.g., Concordia in Canada).
In Section ~\ref{subsec:quality1} we will show that the text of zoombombing posts often further identifies the institution or organization that the zoombombing link belongs to.

\begin{figure}
	\centering
	\includegraphics[width=0.7\linewidth]{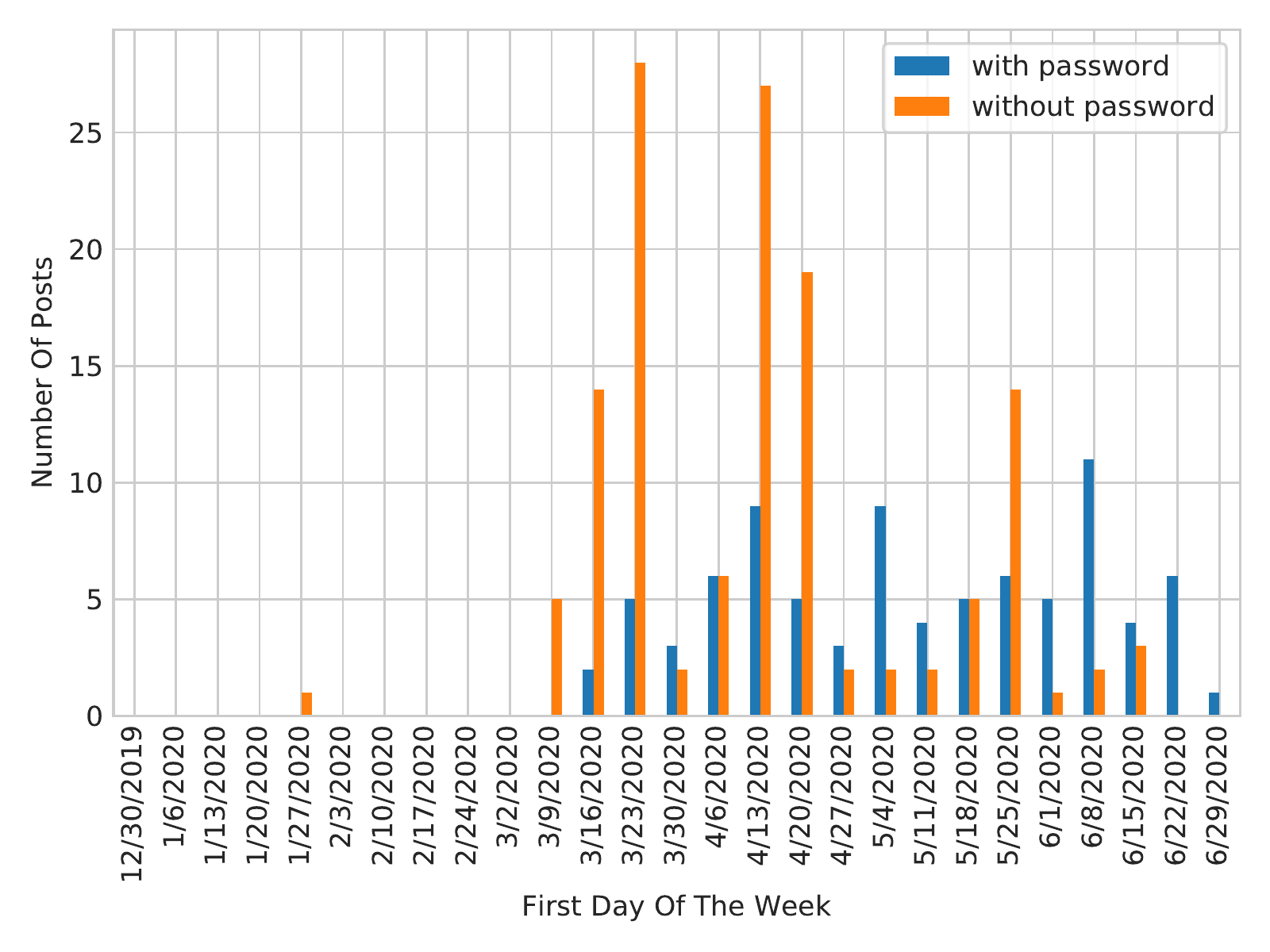}
	\caption{Occurrences of zoombombing links with and without passwords.}
	\label{fig:pwdlink}
\end{figure}

\noindent\textbf{Password protection.} As we discussed in Section~\ref{subsec:meetingservices}, two of the ten online meeting services (Zoom and Webex) allow hosts to protect their meetings using passwords.
In the case of Zoom, the password can be embedded in links as a URL parameter (for example \url{https://zoom.us/j/123456789?pwd=12345aAbBcC678}).
We find that 20 out of the 123 bombing invitations on 4chan, and 64 out of the 95 ones on Twitter include a password.
This is interesting, because the password option was added by Zoom after the quarantine started to curb zoombombing.
In fact, we find that zoombombing posts containing passwords are concentrated toward the latter part of our timeline (see Figure~\ref{fig:pwdlink}).
This is a worrying trend, since as we will confirm in Section~\ref{subsec:quality1} it is an indication that many attacks are called for by insiders who have legitimate access to the meetings, questioning existing security measures and calling for rethinking them.

\subsection{Content Analysis}

\begin{table}[]
\centering
  \scalebox{0.8}{
\begin{tabular}{ll|ll|ll|ll}
\multicolumn{4}{c|}{\textbf{Bombing}}                                                                                                         & \multicolumn{4}{c}{\textbf{Non Bombing}}                                                                                                     \\ \hline
\multicolumn{2}{c|}{\textbf{4chan}}                                   & \multicolumn{2}{c|}{\textbf{Twitter}}                                 & \multicolumn{2}{c|}{\textbf{4chan}}                                   & \multicolumn{2}{c}{\textbf{Twitter}}                                  \\ \hline
\multirow{2}{*}{\textbf{Word}} & \multirow{2}{*}{\textbf{Sim.}} & \multirow{2}{*}{\textbf{Word}} & \multirow{2}{*}{\textbf{Sim.}} & \multirow{2}{*}{\textbf{Word}} & \multirow{2}{*}{\textbf{Sim.}} & \multirow{2}{*}{\textbf{Word}} & \multirow{2}{*}{\textbf{Sim.}} \\
                               &                                      &                                &                                      &                                &                                      &                                &                                      \\ \hline
virtual                        & 0.834                                & zoomcodes                      & 0.860                                & nihilist                       & 0.628                                & live                           & 0.264                                \\
lecture                        & 0.820                                & boys                           & 0.819                                & cia                           & 0.561                                & virtual                           & 0.249                                \\
lesson                     & 0.777                                & zoin                           & 0.814                                & join                            & 0.552                                & pm                             & 0.247                                \\
class                          & 0.774                                & zoomclasse                     & 0.812                                & neo                           & 0.549                                & zoom                        & 0.239                                \\
crash                         & 0.755                                & girls                          & 0.802                                & program                            & 0.505                                & link                           & 0.239                                \\
join                        & 0.697                                & pm                             & 0.792                                & nazi                        & 0.502                                & join                        & 0.229                                \\
webex                          & 0.685                                & raiding                        & 0.785                                & goat                           & 0.482                                & please                          & 0.208                                \\
meeting                           & 0.682                                & random                         & 0.771                                & glownigger                     & 0.478                                & detail                           & 0.195                                \\
conference                         & 0.681                                & shit                           & 0.771                                & fbi                            & 0.455                                & march                        & 0.192                                \\
password                       & 0.675                                & join                           & 0.769                                & autistic                    & 0.374                                & reminder                           & 0.178                               
\end{tabular}
  }
  \caption{Top 10 most similar words (by cosine similarity) related to online meeting links in Bombing \& Non Bombing Threads and Tweets.}
\label{tab:mostsimilar}
\end{table}

After looking at timing information and at the characteristics of URLs, we focus on analyzing the language of social media posts containing zoombombing invitations on Twitter and 4chan, together with their threads on 4chan.
To this end, we leverage \emph{word embedding} models (i.e., word2vec~\cite{mikolov2013efficient}) to quantitatively learn about the context in which zoombombing links are discussed. 
Intuitively, this allows us to identify common themes used in discussions where the links appear.
To build our models, we first replace any meeting link with a keyword ``meetinglink.''

For both 4chan and Twitter, we trained two word2vec models, one for posts (and threads in the case of 4chan) containing zoombombing links, and one for posts and threads containing benign meeting links.
On 4chan, we used a window size of 7 by taking into consideration words that appear at least 5 and 84 times, respectively, for bombing and non-bombing threads, maintaining the ratio of the total amount of posts left after preprocessing. 
To avoid the effect of common/unnecessary words in our model, we removed stop words, punctuation, other URLs, mentions, posts with only one word, and exact quotes of previous posts in the case of threads. 
We also lemmatized the posts and converted all text to lowercase, to avoid weakening the influence of words that are actually the same words or inflected form of the same word.   
On Twitter we applied the same pre-processing techniques as 4chan, and in addition we removed emojis, numbers, and some Twitter-related keywords like RT and FAV while also removing non-alphanumeric characters from words.
Since tweets are usually shorter than 4chan posts, to build our word2vec models we used a window size of 5.
We keep words that appear at least 7 times for non-bombing tweets and words that appear at least once for bombing tweets considering the same ratio we applied for 4chan.

Since online meeting links do not have a fixed position in posts, but attackers place them arbitrarily as a word inside of a sentence, we decided to use the Continuous Bag-Of-Words Model (CBOW)\cite{mikolov2013efficient} for training our word2vec models. 

\noindent\textbf{Most representative words.}
After building our models, we want to identify the words that are closer to zoombombing and non-bombing links on both Twitter and 4chan.
To do this, we looked for the most similar words to 'meetinglink' with respect to the cosine similarities of the vector embeddings of words in our trained models. 

As seen in Table~\ref{tab:mostsimilar}, the most representative words for zoombombing and non-bombing content are very different.
On 4chan, we notice that most zoombombing words are related to education (e.g., ``lecture,'' ``class'') or business meetings (e.g., ``meeting,'' ``conference'').
On Twitter, we observe references to education as well (``zoomclass'') as well as keywords related to attacks (e.g., ``raiding'').
For non-bombing content, on Twitter we can observe that most keywords are related to conference meetings, and reflect the fact that public meeting URLs are often posted on the platform.
On 4chan, we observe that non-bombing meeting URLs are often related to trolling and political discussion. 

\noindent\textbf{Visualizing discussion themes.}
We next aim to identify recurring ``themes'' in zoombombing content.
To this end, we visualize the relationship between the words related to online meeting links following the methodology of Zannettou et al.~\cite{DBLP:journals/corr/abs-1809-01644}. 
From the word2vec models we trained, we create a two-hop ego network around ``meetinglink'' where words are nodes, and the edges are weighted with the cosine similarity between the word embedding vectors of those words; we keep any edge whose weight is greater than or equal to a pre-defined threshold, and visualize this as a graph. 
For each graph, we elect the threshold as the value that results in a graph with 100 nodes (for ease of representation). 
We then detect ``communities'' of words using the Louvain algorithm~\cite{Blondel_2008}, and display them using Gephi's ForceAtlas2 algorithm~\cite{10.1371/journal.pone.0098679}.

\begin{figure*}
	\centering
	\begin{minipage}[c]{\columnwidth}
	\includegraphics[width=\textwidth]{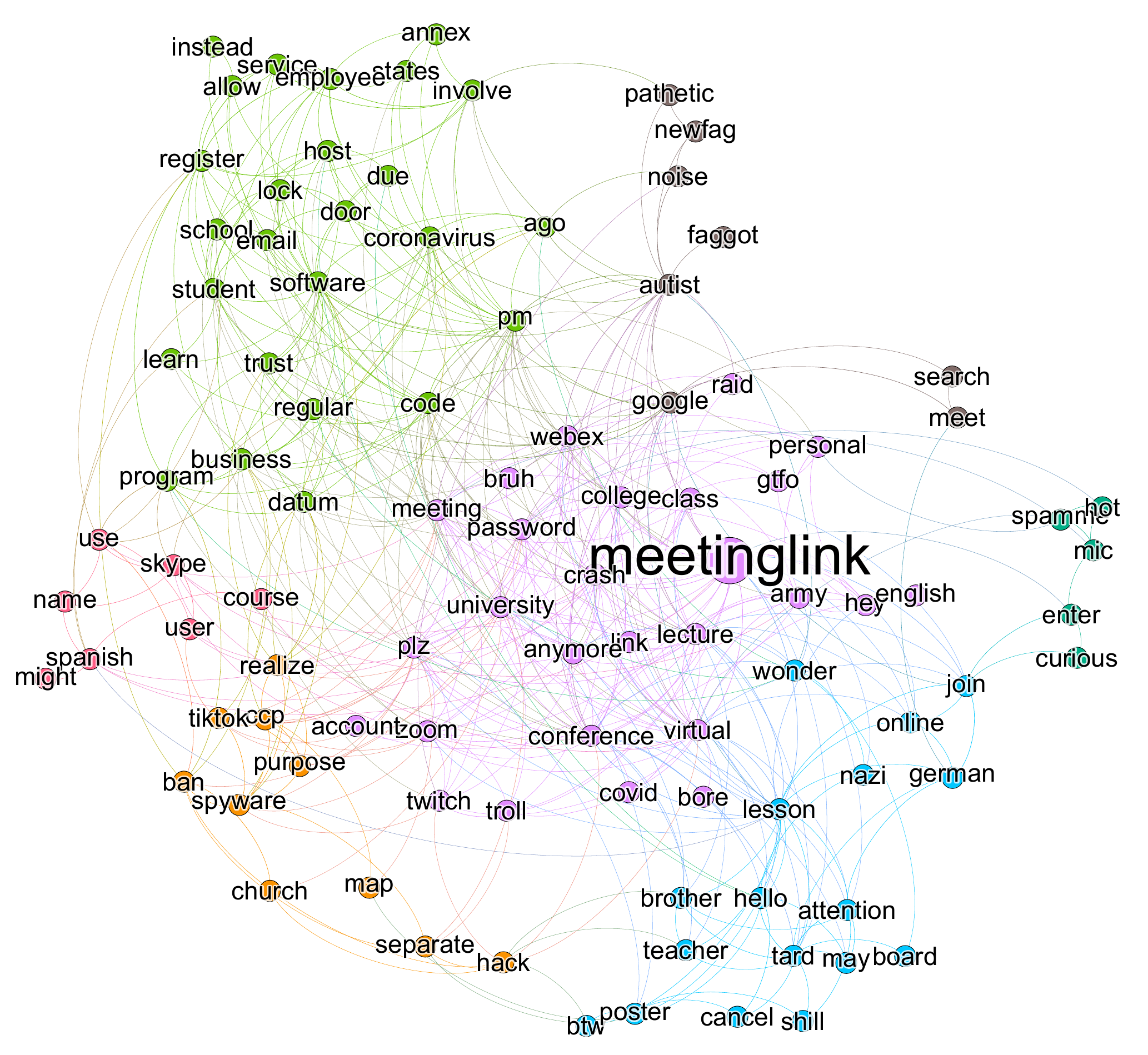}
	\caption{ Words and themes associated with zoombombing links on 4chan.}
	\label{fig:zb_cbow}
	\end{minipage}
    \begin{minipage}[c]{\columnwidth}
	\includegraphics[width=\textwidth]{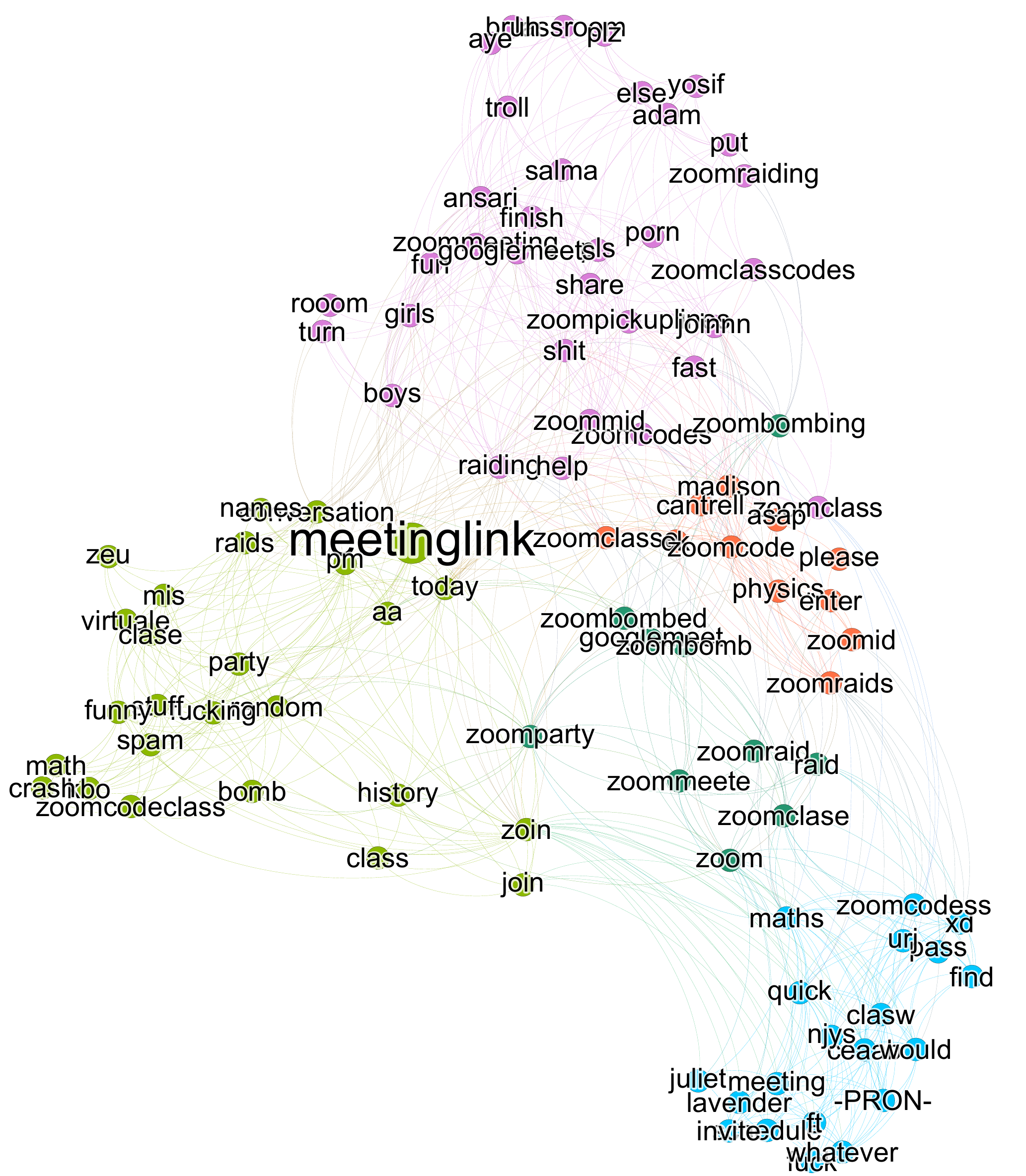}
	\caption{ Words and themes associated with zoombombing links on Twitter.}
	\label{fig:zb_tw}
	\end{minipage}
\end{figure*}

Figures~\ref{fig:zb_cbow} and~\ref{fig:zb_tw} show the results of this analysis for zoombombing invitations in 4chan threads and Twitter posts, respectively.
Intuitively, each colored community can be interpreted as a ``theme'' that is featured prominently in these posts. 
Looking at the 4chan graph (Figure~\ref{fig:zb_cbow}) we can see that many of the themes feature educational topics (e.g., the red community with ``spanish,'' ``course,'' and ``skype'' and the purple community with ``university,'' ``college,'' and ``class'').
We can also see a community (orange) where users talk about security issues/conspiracies as we can derive from words like ``ccp,'' ``tiktok,'' ``spyware,'' and ``ban.''
This indicates that conspiratorial content is not only commonplace in regular discussion on 4chan, but is also featured in zoombombing content.
See the following post for example:

\begin{mdframed}[style=MyFrame,nobreak=true,align=center]

\begin{quote}
``If you do the research you'll see our MSM is in bed with the CCP. This is being utilized for propaganda purposes just like tiktok. I work with a bunch of regressed and they all love posting on tiktok. The users of these applications have close to zero foresight when it comes to Intel collection in any fashion from any party. Kind of we are fucked because Jews take chinese money as investments in their companies.''
\end{quote}

\end{mdframed}

On Twitter (Figure~\ref{fig:zb_tw}) we can again see themes that cover online classes (e.g., the green community with ``class,'' ``history,'' ``math'').
We also see a number of keywords that are used as hashtags on the platform to ensure that the calls for zoombombing obtain more visibility (e.g., ``zoomcodeclass,'' ``zoombomb,'' ``zoomraids'').

\begin{figure*}
	\centering
	\begin{minipage}[c]{\columnwidth}
	\includegraphics[width=\textwidth]{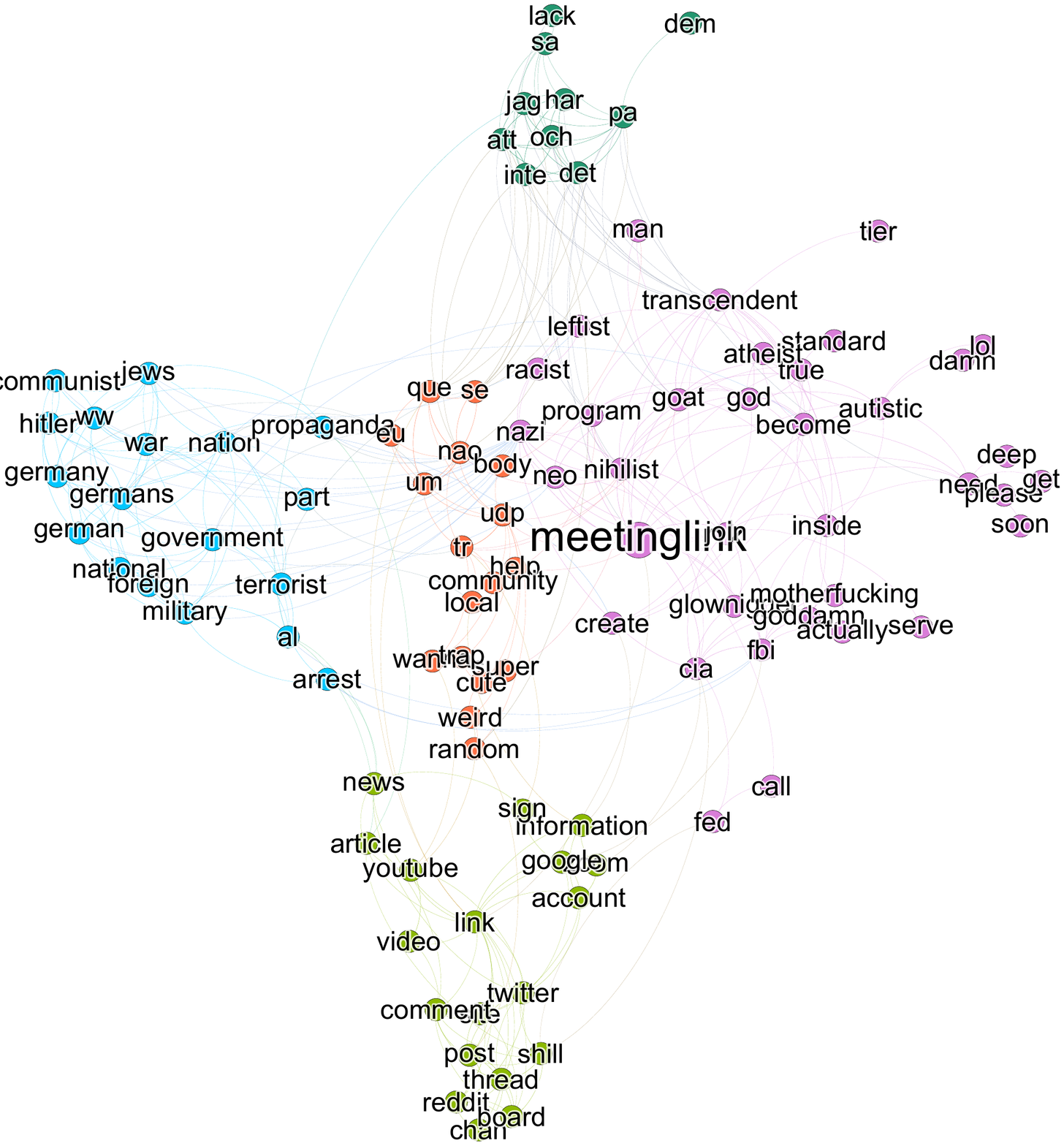}
	\caption{ Words and themes associated with online meeting links on non-bombing threads on 4chan.}
	\label{fig:nzb_cbow}
	\end{minipage}
	\begin{minipage}[c]{\columnwidth}
	\includegraphics[width=\textwidth]{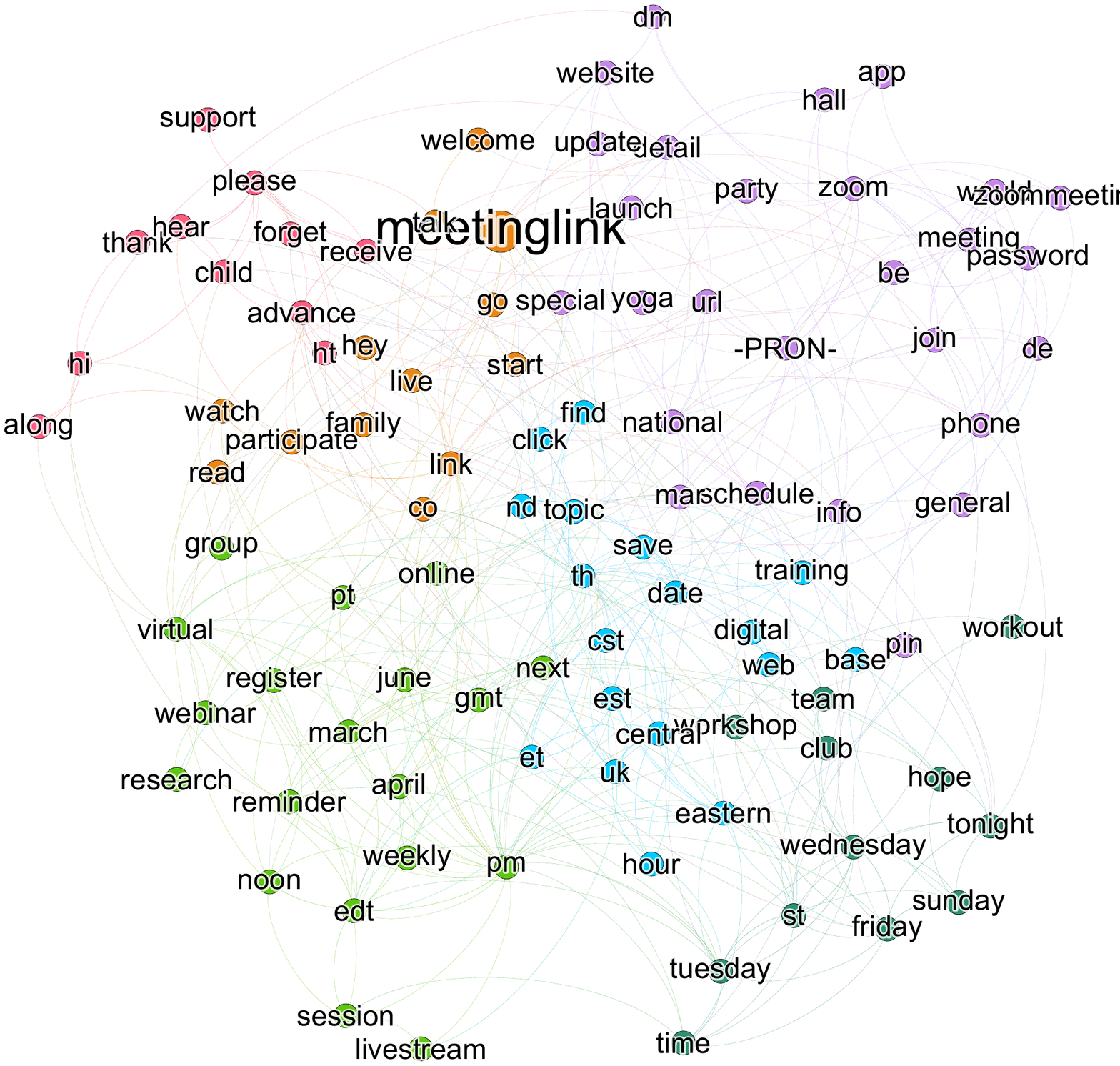}
	\caption{ Words and themes associated with online meeting links on non-bombing tweets on Twitter.}
	\label{fig:nzb_tw}
	\end{minipage}

\end{figure*}

For completeness, Figures~\ref{fig:nzb_cbow} and~\ref{fig:nzb_tw}) show the graphs for non-bombing threads on 4chan and non-bombing tweets on Twitter. 
As it can be seen, the themes in these cases are more varied.

\section{Qualitative Analysis: Understanding  Forum Content}

Our quantitative analysis highlighted several interesting aspects of zoombombing invitations and their discussion.
In particular, we found evidence that online classes in particular are targeted by attacks, and we found several meeting passwords included in invitations, which could be an indicator that attacks are called for by insiders who have legitimate access to the meeting rooms.
When dealing with online activity carried out by humans, however, quantitative analysis can only identify general trends, and lacks the nuance required to provide a better understanding of the problem.
In this section, we answer deeper questions via a more thorough qualitative analysis informed by our quantitative results.
As explained in Section \ref{section:annotation} this analysis was conducted by having four authors of the paper manually annotate the dataset.
Where appropriate, our analysis covers zoombombing posts on Twitter and 4chan, while for some of the analysis (for example the one analyzing back and forth communication between attackers) we only rely on 4chan threads.
Based on our threat model (see Section~\ref{ThreatModel}), we analyze attacks across four phases: i)~\emph{Call for attack}, ii)~\emph{Coordination}, iii)~\emph{Delivery}, and iv)~\emph{Harm}.

\subsection{Phase I: Call for attack }
\label{subsec:quality1}

In this phase, an attacker posts a call for an attack on an online platform.

\descr{Targeting the class room.}
In Section \ref{subsec:characterzblink} we showed that we could quantitatively identify 8 academic institutions targeted by zoombombing attacks on 4chan.
In addition to information that can be directly extracted from the URL of the bombing link, many bombing posts include additional text indicating that online classes are the target.
For example, ``lecture,'' ``teacher,'' ``class,'' etc. show up regularly in these threads.
We find that 91 of our 123 zoombombing threads on 4chan target online classes.
Of the 32 remaining threads, three target business meetings, and the target of the remainder could not be conclusively determined. 
On Twitter, we find that 56 of our 95 bombing calls target schools. 

\descr{Evidence of insiders' complicity.}
In Section~\ref{subsec:characterzblink}  we showed that 11 zoombombing links on 4chan included passwords, indicating that who called for the attack was a legitimate participant in the meeting (e.g., a student in the class). 
When annotating the threads, we find 9 additional zoombombing threads including a password in the body of messages. 
In total, this accounts for 20 of our 123 threads on 4chan.
For Twitter, we showed that 64 out of the 95 tweets included a password in the zoombombing link.

There are additional indicators that can be used to qualitatively determine if an attack is called by an insider.
In this section, we look for two indicators: 1)~whether the language of the call of the attack suggests that the attack is called by an insider and 2)~whether whoever calls for an attack shares knowledge about the meeting that only an insider would have.

For the first aspect, we look for language like ``my lecture,'' ``my colleague's presentation,'' ``my company's meeting,'' etc.
58 zoombombing threads on 4chan  and 19 zoombombing tweets on Twitter include such language indicating that the attack is called by an insider.
In many cases, the users calling for the attack provide additional information that only an insider would know.
In 8 zoombombing threads and 8 zoombombing tweets, the attacker asks others to use a certain name when joining the meeting to avoid being identified as an intruder and removed.

\begin{mdframed}[style=MyFrame,nobreak=true,align=center]
\begin{quote}
``[GOOGLEMEETURL] name yourself ``WONG SHIU PING TONY'' all caps or she wont let you in.''\\
``Also please use real-sounding names.''
\end{quote}
\end{mdframed}

In 11 threads we learn that the attacker is an insider from their interaction with other users.
\begin{mdframed}[style=MyFrame,nobreak=true,align=center]
\begin{quote}
``Same school as you, different major. Someone wrote "NIGGERS" in my zoom class with the annotate function and started a zoom fight.''
\end{quote}
\end{mdframed}

Together with all information from both meeting links and post text, we identify 86 out of 123 zoombombing threads on 4chan that appear to have been posted by insiders (38/54 for Zoom, 35/46 for Google Meet, 8/10 for Cisco Webex, 3/3 for Skype, 0/2 for GoToMeeting, 2/7 for Jitsi, and 0/1 Teams).
For Twitter, we find that 78 out of the 95 zoombombing tweets were posted by insiders. 

\descr{Failed calls to attack.}
While 100 (out of 123) of our threads did start with an invitation to bomb, 46 (out of 100) of these received no further replies.
I.e., the call for an attack seems to have been stillborn.
For the threads with replies, 54 (out of 77) were started with an invitation to bomb and 23 (out of 77) were created with more general topics of interest (e.g., politics, COVID-19, etc.) which were later converted into bombing threads.
Threads with general topics tend to attract more posts than bombing threads.

\subsection{Phase II: Coordination}
\label{subsec:quality2}

After posting an invite to a zoombombing, attackers coordinate to carry it out.
To better understand this, we look for temporal information on when the attack should be carried out in both 4chan threads and tweets.

\descr{Crimes of opportunity.}
Considering that most of the zoombombing links target online classes, and that these occur at regularly scheduled times, there is a question as to how much premeditation goes into a bombing attack.
On the surface, it seems plausible that attacks could be planned days, and even weeks in advance.
To dig deeper, we looked at the text posted along with a link and determined whether or not the invite was for a live meeting, or one that was scheduled to take place in the future.
I.e., are attackers asking people to bomb \emph{right now} or planning a bombing that is going to happen later?
We found that 115 of 123 bombing links on 4chan and 93 of 95 links on Twitter came along with a clear implication that the meeting was live at the time of posting.
We find 8 future links among 123 links on 4chan and 2 out of 95 links on Twitter.
A future link example from 4chan is:

\begin{mdframed}[style=MyFrame,nobreak=true,align=center]

\begin{quote}
  ``RAID THIS BOOMER Wednesdays 10:00-10:45 [INSTITUTIONAL ZOOMURL]''  
\end{quote}
\end{mdframed}

\descr{Refusing to participate.}
We find 20 threads on 4chan where users openly refuse to join into the attack, calling it unethical or referring to the fact that 4chan users are not the insider's personal army (NYPA -- Not Your Personal Army).
This indicates that not all users on 4chan are willing to participate in these attacks, and is particularly interesting because it is a possible explanation for at least some failed attacks: users do not reply because they reject the idea of being a troll in the service of another user.

\begin{mdframed}[style=MyFrame,nobreak=true,align=center]
\begin{quote}
``[ZOOMURL]please spam this online class''\\
``I'm not downloading shit''\\ 
``Nypa faggot''
\end{quote}
\end{mdframed}

\subsection{Phase III: Delivery}

In this phase, the attackers join the online meeting and begin their harassing and disruptive actions.
As part of our analysis, we find discussion of how the attacks went down in replies within the bombing threads on 4chan.

\descr{Quick action.}
We compare the time interval between when the link is posted and the first feedback on the attack.
Of 123 bombing threads on 4chan, we find 37 with clear feedback related to the bombing.
According to this analysis, a zoombombing attack finishes within 20 minutes.
An example of attack feedback on 4chan is as follows:

\begin{mdframed}[style=MyFrame,nobreak=true,align=center]
\begin{quote}
19:51:59
``Join a teachers zoom [ZOOMURL]''\\
20:05:18
``What the fuck is this? Who are these people?''\\
20:07:43
``quickly screencap it. They kicked me out instantly.''
\end{quote}
\end{mdframed}

\descr{Problem feedback.}
For 24 threads we find participants reporting problems with the zoombombing invitation.

\begin{mdframed}[style=MyFrame,nobreak=true,align=center]
\begin{quote}
``Raid our school live call class, i believe in you faggots. [GOOGLEMEETLINK]''\\
``It says someone has to allow me to join, some shit like that''\\
``this meeting has been locked by the host. Sad!''
\end{quote}
\end{mdframed}

\subsection{Phase IV: Harm}

Finally, we want to understand the toxic speech that happens during attacks, together with what actions attackers carry out.

\descr{Toxic speech.}
We find 14 4chan zoombimbombing threads containing toxic content including racism, sexism, or hateful words.

\begin{mdframed}[style=MyFrame,nobreak=true,align=center]
\begin{quote}
``[SKYPEURL] Anyone wanna join our online lesson? Our teacher is black. Its gonna be in 20 mins.''\\
``NIGGER.''
``That is absolutely a ‘he’, no matter how the swine identifies.''\\
``What the fuck, I swear I spotted a beard on that chin.''
\end{quote}
\end{mdframed}

On Twitter, we did not find any toxic tweets among the 95 zoombombing tweets.
However, recall that on Twitter we only retrieved the call for attacks and do not have any feedback (e.g., the replies to those tweets). 

\descr{Crime scene feedback.}
On 4chan, we find 15 threads containing feedback from the zoombombing attack, providing us with a better view of what happens during these attacks.
Here are some examples:
\begin{mdframed}[style=MyFrame,nobreak=true,align=center]
\begin{quote}
``Hard working he's probably the kind of teacher who sits reverse on a chair and is up to date with the cool kids.''\\
``HAHAHAHA that was great.''\\
``Party's over my dudes, IT is here shutting down the stream, we had a good laugh.''\\
``Did you hear me saying nigger?''\\
``Ayone heard me farting.''\\
``Yeah everyone heard and saw the chat and vc lmao.''\\
``I didn't hear that, maybe not loud enough but there was a bunch of rambling about the numbers on screen and then someone started farting and the class was just dying of laughter.''\\
``Nice bro.''\\
`Totally lmfao. Best class disruption ever.''
\end{quote}
\end{mdframed}

\section{Discussion}

In this paper we have presented a data-driven analysis of the emerging phenomenon of zoombombing.
Our findings improve the understanding of who the people calling for zoombombing attacks are and how they operate.
In the following, we first discuss the implications of our findings to existing mitigations against zoombombing, and propose some best practices to protect online meeting rooms.
We then discuss the limitations of our study and some future work directions.

\descr{Implications for zoombombing mitigation.}
After the rise in popularity of online meeting tools, researchers have been looking at the privacy risks linked to online meeting~\cite{kagan2020zooming}.
At the same time, researchers, law enforcement, and the online meeting providers themselves have been publishing best practices to avoid zoombombing~\cite{brown2020notes,fbi,zoomlog_keep}.
These include not posting meeting links publicly, protecting meeting rooms to control who can get in, and reducing the capabilities of participants, like muting them upon joining and disabling screen sharing and screen annotations.

The main assumption behind existing guidelines to prevent zoombombing is that attackers will find meeting links online, or that they will bruteforce their ID.
Given this threat model, protecting meetings with passwords makes sense.
However, our findings show that most of the calls for attacks we observe come from insiders.
This makes password protection ineffective, because the insider will share the password with the other attackers.
Having participants join a waiting room and vet them before letting them in can be a more effective mitigation, although it inevitably increases the workload of meeting hosts, requiring moderators specifically checking the meeting room in the case of large meetings.
Our analysis however shows that insiders often share additional information with potential attackers, for example instructing them to select names that correspond to legitimate participants in the meeting.
This reduces the effectiveness of a waiting room, because it makes it more difficult for hosts and moderators to identify intruders.

Providing a unique link for each participant reduces the chances of success of zoombombing attacks.
If the meeting service still allows multiple people joining with the same link, at least this gives some accountability, since the meeting host can identify who the insider was based on the unique link used by attackers to join.
An even better mitigation is to allow each participant to join using a personalized meeting link.
This way, as long as the insider joins the meeting unauthorized people will not be able to join using the same link.
While this mitigation makes zoombombing unfeasible, not all meeting services have adopted it. 
At the moment of writing, only Zoom and Webex allow per-participant links that allow a single user to join at a time.
To do this, Zoom requires participants to log in, and checks if the unique link is the same that was sent to that email address as a calendar invite.
We encourage other meeting platforms to adopt similar access control measures to protect their meetings from insider threats.

Additionally, we find that zoombombing attacks usually happen in an opportunistic fashion, with insiders asking others to join meeting happening in real time.
This reduces the effectiveness of proactive measures like monitoring social media for calls for future attacks.

\descr{Limitations and future work.}
As any data-driven study, our study is not exempt from limitations.
We only have a 1\% sample of Twitter available, therefore our zoombombing results on the platform are a lower bound of the actual extent on the problem.
Additionally, the API limitations prevent us from collecting replies to the zoombombing tweets, allowing us to only get a partial picture of how attacks unfold on the platform.
On 4chan, users are anonymous.
We therefore cannot trace per-user behavior, and this prevents us from observing serial offenders calling for multiple attacks over time.
Finally, our analysis is limited to calls for attacks and responses to such calls on social media, but we are unable to observe what happens in the actual meeting rooms.
Future work could develop alternative study designs that allow analyzing the attack on the online meeting platform itself, for example by collecting and analyzing recorded online meetings that were bombed, or by interviewing victims of zoombombing.
This would also allow a better understanding of the mental and emotional toll that zoombombing victims have to go through.

\section{Related Work}

\noindent\textbf{Coordinated malicious activity on social media.}
The security community has extensively studied automated malicious behavior on social media, mostly focusing on bots sending spam~\cite{gao2010detecting,grier2010spam,yuan2019detecting} and on malicious accounts colluding to inflate each other's reputation~\cite{de2014paying,stringhini2013follow,weerasinghe2020pod}.
The mitigation systems proposed to detect and block this type of activity rely on the fact that these operations are large scale, rely on automated methods, and are carried out by single entities.
Therefore, synchronization features can be used to distinguish between benign and malicious activity~\cite{cao2014uncovering,stringhini2015evilcohort,zhao2009botgraph}.
Alternatively, systems have been proposed that identify common traits in massively created fake accounts, for example an anomalous fraction of followers to friends or a large set of accounts created around the same time~\cite{benevenuto2010detecting,davis2016botornot,stringhini2010detecting,yang2011free,yuan2019detecting}.

More recently, the community's focus expanded to looking at coordinated malicious campaigns that are not carried out by automated means, but rather by humans controlling a small number of inauthentic accounts.
This includes conspiracy theories being pushed on social media~\cite{starbird2017examining,starbird2019disinformation} and influence campaigns by foreign state actors~\cite{badawy2018analyzing,zannettou2019let}.
While not as automated as large-scale bot activity, these campaigns still show coordination, which can be leveraged for detection~\cite{luceri2020detecting}.

\noindent\textbf{Coordinated online harassment and aggression.}
A closer line of work to the problem studied in this paper looks at coordinated behavior geared toward harassing victims online.
Kumar et al.~\cite{kumar2018community} measure the problem of \emph{brigading} on Reddit, where the members of one sub-community (\emph{subreddit}) organize to disrupt another community by posting offensive messages and prevent it from continuing its normal operation.

Hine et al.~\cite{hine2017kek} study the activity of 4chan's Politically Incorrect Board (/pol/), showing that members of that community often call for attacks against people who posted videos on YouTube, ending up harassing the poster in the comments section of the video.
Mariconti et al.~\cite{mariconti2019you} develop a multi-modal machine learning system able to predict which videos are likely to receive this kind of hate attacks, in the hope of aiding moderation efforts.

Zannettou et al.~\cite{zannettou2020measuring} investigate a similar phenomenon, studying the effect of posting a URL to a news article on 4chan and Reddit.
They show that posting URLs to certain types of news outlets results in a sudden increase in the hate speech on the comments to that article.

Snyder et al.~\cite{snyder2017fifteen} study the problem of \emph{doxing}, in which attackers post information about a victim, calling for people to attack that person through multiple media (e.g., on multiple social networks or through email), sometimes even transcending to the physical world.

Tseng et al.~\cite{tseng2020tools} analyze five forums in which miscreants share and discuss tools and techniques that can be used to spy on their partners and further harass them.

Our work builds on previous research on coordinated harassment by studying the emerging problem of zoombombing. 
Unlike previously studied threats, we show that zoombombing attacks are often called by insiders; this has important implications when designing security mitigations against the problem.

\section{Conclusion}
In this paper, we performed the first data-driven study of calls for zoombombing attacks on social media.
Our findings indicate that these attacks mostly target online lectures, and that they are mostly called by insiders who have legitimate access to the meetings.
We find that insiders are commonly sharing confidential information like meeting passwords and the identify of real participants in the meeting, making common protections against zoombombing ineffective.
We also find that calls for zoombombing are usually targeting meetings happening in real time, making the proactive identification of such attacks challenging.
To protect against the threat, we encourage online meeting services to allow hosts to create unique meeting links for each participant, although we acknowledge that this has usability implications and might not always be feasible.

\bibliographystyle{IEEEtran}
\bibliography{refs.bib}

\end{document}